\documentclass[english,A4,superscriptaddress,nofootinbib,prd]{revtex4}
\usepackage[OT1]{fontenc}
\usepackage[latin9]{inputenc}
\usepackage{xcolor}
\usepackage{babel}
\usepackage{amsbsy}
\usepackage{amstext}
\usepackage{amssymb}
\usepackage{esint}
\usepackage[unicode=true,pdfusetitle,
 bookmarks=true,bookmarksnumbered=true,bookmarksopen=true,bookmarksopenlevel=1,
 breaklinks=false,pdfborder={0 0 1},backref=false,colorlinks=true]
 {hyperref}
\hypersetup{
 linkcolor=blue,citecolor=cyan}

\makeatletter
\@ifundefined{textcolor}{}
{%
 \definecolor{BLACK}{gray}{0}
 \definecolor{WHITE}{gray}{1}
 \definecolor{RED}{rgb}{1,0,0}
 \definecolor{GREEN}{rgb}{0,1,0}
 \definecolor{BLUE}{rgb}{0,0,1}
 \definecolor{CYAN}{cmyk}{1,0,0,0}
 \definecolor{MAGENTA}{cmyk}{0,1,0,0}
 \definecolor{YELLOW}{cmyk}{0,0,1,0}
}

\usepackage{babel}

\makeatother

\begin{document}
\title{Corrections from space-time dependent electromagnetic fields\\
to Wigner functions and spin polarization}
\author{Shi-Zheng Yang}
\email{yangsz@ustc.edu.cn}
\affiliation{Department of Modern Physics, University of Science and Technology
of China, Anhui 230026, China}
\author{Jian-Hua Gao}
\email{gaojh@sdu.edu.cn}
\affiliation{Shandong Key Laboratory of Optical Astronomy and Solar-Terrestrial
Environment, School of Space Science and Physics, Shandong University,
Weihai, Shandong 264209, China}
\author{Shi Pu}
\email{shipu@ustc.edu.cn}
\affiliation{Department of Modern Physics, University of Science and Technology
of China, Anhui 230026, China}
\affiliation{Southern Center for Nuclear-Science Theory (SCNT), Institute of Modern
Physics, Chinese Academy of Sciences, Huizhou 516000, Guangdong Province,
China}
\begin{abstract}
We have derived the Wigner equations at global equilibrium with constant vorticity but space-time dependent electromagnetic fields up to second order in semiclassical expansion.
We obtain the new second-order contributions to the charge currents and energy-momentum tensor from the varying electromagnetic fields.
We also compute the new corrections to the spin polarization pesudo-vector from both contant and varying electromagnetic fields.
We also find that the space-time dependent electromagnetic field provides a tighter constraint on the solutions of Wigner functions in global
equilibrium compared with constant electromagnetic field.
\end{abstract}
\maketitle

\section{introduction\label{sec:introduction}}

Relativistic heavy ion collision provides a unique opportunity to
generate huge vorticity \citep{Liang:2004ph,Gao:2007bc,Becattini:2007sr}
and electromagnetic (EM) fields \citep{Bzdak:2011yy,Deng:2012pc,Bloczynski:2012en},
which can lead to several novel quantum effects such as chiral magnetic
effect (CME) \citep{Vilenkin:1980fu,Kharzeev:2007jp,Fukushima:2008xe}
and chiral vortical effect (CVE) \citep{Vilenkin:1978hb,Kharzeev:2007tn,Erdmenger:2008rm,Banerjee:2008th}.
{The CME and CVE are the generation of the vector current induced by a magnetic and vortical
field in the presence of chirality imbalance, respectively. Besides CME and CVE, }
the axial current can also be generated by magnetic field and vorticity,
namely chiral separation effect (CSE) \citep{Son:2004tq,Metlitski:2005pr}
and local polarization effect (LPE) \citep{Gao:2012ix}. These chiral
transports have been extensively studied within various approaches such
as AdS/CFT \citep{Newman:2005hd,Yee:2009vw,Rebhan:2009vc,Gorsky:2010xu,Gynther:2010ed,Hoyos:2011us,Amado:2011zx,Nair:2011mk,Kalaydzhyan:2011vx,Lin:2013sga,Pu:2014cwa,Pu:2014fva},
quantum field theory \citep{Kharzeev:2007jp,Fukushima:2008xe,Kharzeev:2009pj,Fukushima:2009ft,Asakawa:2010bu,Fukushima:2010vw,Fukushima:2010zza,Landsteiner:2011cp,Hou:2011ze,Hou:2012xg,Chen:2016xtg,Lin:2018aon,Feng:2018tpb,Dong:2020zci},
relativistic hydrodynamic \citep{Son:2009tf,Sadofyev:2010pr,Pu:2010as,Kharzeev:2011ds},
and chiral kinetic theory \citep{Gao:2012ix,Stephanov:2012ki,Son:2012zy,Chen:2012ca,Manuel:2013zaa,Chen:2014cla,Chen:2015gta,Hidaka:2016yjf,Mueller:2017lzw,Huang:2017tsq,Huang:2018wdl,Hidaka:2018ekt,Gao:2018wmr,Gao:2018jsi,Liu:2018xip}.
{On the other hand, the spin polarization of $\Lambda$ and $\overline{\Lambda}$ hyperons \citep{STAR:2017ckg,STAR:2019erd,ALICE:2019aid,STAR:2020xbm} and spin alignment of vector mesons \citep{STAR:2022fan,ALICE:2019aid,ALICE:2023jad}, proposed by Refs.~\citep{Liang:2004ph,Liang:2004xn,Gao:2007bc}, have been measured very recently.
The global polarization can be well described by various phenological models \citep{Karpenko:2016jyx,Becattini:2017gcx,Xie:2017upb,Pang:2016igs,Li:2017slc,Wei:2018zfb,Ryu:2021lnx,Shi:2017wpk,Fu:2020oxj,Sun:2017xhx,Wu:2022mkr,Alzhrani:2022dpi} under the assumption of the global equilibrium. The local polarization requires the effects beyond the global equilibrium, such as the polarization induced by the shear tensor and the gradient of chemical potential over temperature \citep{Liu:2021uhn,Fu:2021pok,Becattini:2021suc,Becattini:2021iol,Hidaka:2017auj,Yi:2021ryh,Fu:2022myl,Wu:2022mkr} and corrections from collisions \citep{Weickgenannt:2020aaf,Yang:2020hri,Weickgenannt:2021cuo,Sheng:2021kfc,Wang:2020pej,Wang:2021qnt,Weickgenannt:2022jes,Lin:2021mvw,Fang:2022ttm,Wang:2022yli,Lin:2022tma,Li:2019qkf,Wagner:2022amr,Wang:2022yli,Lin:2022tma,Yamamoto:2023okm,Fang:2023bbw,Lin:2024zik,Fang:2024vds}. However, polarization along the beam direction in high transverse momentum or low multiplicity
regions in relativistic nucleus-nucleus collisions \citep{STAR:2023eck},
and in p-Pb collisions \citep{Yi:2024kwu} are still open questions. For more details, one can see the recent reviews \citep{Becattini:2020ngo,Gao:2020vbh,Hidaka:2022dmn,Becattini:2022zvf,Becattini:2024uha} and the references therein.
}

The EM field in non-central heavy ion collisions can be not
only very strong but also highly dependent on space and time.
It turns out that the EM field could decrease one order of magnitude within 0.2 fm/c
at the beginning of the collision and vary steeply within 10 fm \citep{Bzdak:2011yy,Deng:2012pc}
in spatial distribution, also see the recent developments from relativistic magnetohydrodyanmics \cite{Pu:2014fva, Roy:2015kma,Roy:2015coa, Pu:2016bxy,Inghirami:2016iru,Pu:2016ayh,Pu:2016rdq,Siddique:2019gqh,Peng:2022cya} and the simulations by kinetic theories \cite{Wang:2021oqq,Yan:2021zjc,Zhang:2022lje}.
Hence it is valuable to study how these vector
and axial currents can be affected by the space-time dependent EM
field. In semi-classical expansion, the CME, CVE, CSE, and LPE are
all first-order chiral effects, and the terms associated with the
derivative of EM field arises at least at second-order of $\hbar$.
The second-order effects have been computed in various approaches,
such as relativistic hydrodynamics \citep{Kharzeev:2011ds}, AdS/CFT
duality \citep{Banerjee:2012iz,Bhattacharyya:2013ida,Megias:2014mba,Bu:2019qmd},
Kubo formula or quantum statistics in quantum field theory \citep{Jimenez-Alba:2015bia,Hattori:2016njk,Buzzegoli:2017cqy,Buzzegoli:2018wpy,Becattini:2020qol,Palermo:2021hlf},
chiral kinetic theory \citep{Satow:2014lia,Gorbar:2017cwv,Gorbar:2017toh,Mameda:2023ueq}.
Most of these works focus on the second-order contribution from the
quadratic terms of vorticity or EM field. Only very few works have
discussed the second-order contribution from the derivative terms
of EM field \citep{Kharzeev:2011ds,Satow:2014lia,Mameda:2023ueq}
when space-time dependent EM field is involved.
{Similar stories occur in the topic of spin polarization. Although the spin polarization induced by magnetic fields has been studied in Refs.~\cite{Muller:2018ibh, Guo:2019joy, Buzzegoli:2022qrr, Xu:2022hql, Peng:2022cya}, the corrections from varying EM fields and the coupling between EM fields and thermal vorticiy up to the
$\mathcal{O}(\hbar^2)$ are still lacking in this field.
}

The Wigner function approach can successfully discribe the first-order
chiral effects, such as CME, CVE, CSE and LPE \citep{Gao:2012ix}
as well as the second-order chiral effects \citep{Yang:2020mtz} and spin polarization induced by the thermal voriticty and shear tensor \citep{Hidaka:2018ekt,Yi:2021ryh,Yi:2021unq, Wu:2022mkr,Yi:2023tgg,Yi:2024kwu} when
the EM field is constant. In this work, we will generalize the Wigner
function approach in Refs.~\citep{Gao:2012ix, Yang:2020mtz,Hidaka:2022dmn} from
constant to space-time varying EM field and find the specific solution
of Wigner equation in global equilibrium with constant vorticity but
varing EM field. The charge currents and energy-momentum tensor can
be directly obtained by integrating the chiral Wigner function over
momentum. {The spin polarization pesudo-vector can also be derived straightforwardly following Refs.~\citep{Yi:2021ryh,Yi:2021unq,Wu:2022mkr,Yi:2023tgg}.}

This paper is organized as follows. In Sec. \ref{sec:Wigner-function-in},
we introduce the Wigner function and equation. In Sec. \ref{sec:Wigner-function-at},
we solve the Wigner equation to obtain chiral Wigner function up to
second-order in global equilibrium. Then we integrate the Wigner function
over momentum to obtain the vector and axial currents in Sec. \ref{sec:Vector-and-axial} and derive the energy-momentum tensor in Sec.\ref{sec:energy-momentum-tensor}. We also compute the corrections from space-time dependent
EM fields to spin polarization pesudo-vector in Sec.~\ref{sec:polarization}.
We discuss the terms which cannot be completely determined by Wigner
equations and present their general expressions in Sec. \ref{sec:general-solution}.
In Sec. \ref{sec:summary}, we summarize our results. Throughout this
work, we choose the metric tensor $g^{\mu\nu}=\mathrm{diag}\{1,-1,-1,-1\}$
and Levi-Civita tensor $\epsilon^{0123}=1$.

\section{Wigner function and equation}
\label{sec:Wigner-function-in} First, we will give a brief review
on the Wigner function formalism. We will focus on chiral fermions
under a background EM field. The Wigner function for spin-1/2 particles
is defined as \citep{Vasak:1987um}
\begin{eqnarray}
W_{\alpha\beta}(x,p) & = & \int\frac{d^{4}y}{(2\pi)^{4}}e^{-ip\cdot y/\hbar}\left\langle \bar{\psi}_{\beta}\left(x+\frac{y}{2}\right)U\left(x+\frac{y}{2},x-\frac{y}{2}\right)\psi_{\alpha}\left(x-\frac{y}{2}\right)\right\rangle ,
\end{eqnarray}
where $\left\langle \cdot\right\rangle $ denotes ensemble average,
$\psi_{\alpha},\bar{\psi}_{\beta}$ are Dirac fields and $U(x_{1},x_{2})=\mathrm{Exp}[-i\int_{x_{2}}^{x_{1}}dz_{\mu}A^{\mu}(z)/\hbar]$
is the gauge link along the straight line from $x_{2}$ to $x_{1}$.
Here $A^{\mu}(x)$ is the vector potential of the background EM field.
The Wigner equation for chiral fermion {without collision interaction} is obtained from Dirac equation
and given by \citep{Vasak:1987um}
\begin{eqnarray}
\gamma_{\mu}(\Pi^{\mu}+\frac{i}{2}\hbar\nabla^{\mu})W(x,p) & = & 0,\label{eq:Wigner_equation_W}
\end{eqnarray}
where $\gamma_{\mu}$ are Dirac matrices and the operators $\Pi^{\mu}$
and $\nabla^{\mu}$ are defined by
\begin{eqnarray}
\Pi^{\mu} & = & p^{\mu}-\frac{1}{2}\hbar j_{1}\left(\frac{1}{2}\hbar\Delta\right)F^{\mu\nu}\partial_{\nu}^{p},\nonumber \\
\nabla^{\mu} & = & \partial^{\mu}-j_{0}\left(\frac{1}{2}\hbar\Delta\right)F^{\mu\nu}\partial_{\nu}^{p}.
\end{eqnarray}
In the definitions above, $j_{0}(z)=\sin z/z$ and $j_{1}(z)=(\sin z-z\cos z)/z^{2}$
are zeroth- and first-order spherical Bessel functions, respectively.
The partial derivative $\partial_{p}^{\mu}(\partial^{\mu})$ denotes
the momentum(space-time) derivative. We have introduced the triangle
operator $\Delta=\partial_{p}\cdot\partial$ with $\partial^{\mu}$
only acting on the $F^{\mu\nu}$ not on the Wigner function. The Wigner
function can be decomposed into the following form
\begin{eqnarray}
W & = & \frac{1}{4}\left[\mathcal{F}+i\gamma^{5}\mathcal{P}+\gamma^{\mu}\mathcal{V}_{\mu}+\gamma^{5}\gamma^{\mu}\mathcal{A}_{\mu}+\frac{1}{2}\sigma^{\mu\nu}\mathcal{S}_{\mu\nu}\right],\label{eq:Clifford_algebra}
\end{eqnarray}
where {the vector and axial components for the Wigner function, $\mathcal{V}^\mu$ and $\mathcal{A}^\mu$, are the vector and axial currents in the phase space. }
For chiral fermions, it is convenient to introduce the chiral Wigner
function defined as
\begin{eqnarray}
\mathcal{J}_{s} & \equiv & \frac{1}{2}(\mathcal{V}_{\mu}+s\mathcal{A}_{\mu}),\label{eq:chiral_Wigner_function}
\end{eqnarray}
where $s=+1$ and $s=-1$ represent the right- and left-handed chiralities,
respectively. Substituting the Eq. (\ref{eq:Clifford_algebra}) into
Eq. (\ref{eq:Wigner_equation_W}) and using the definition (\ref{eq:chiral_Wigner_function}),
we obtain the equations of motion for chiral Wigner function
\begin{eqnarray}
\Pi\cdot\mathcal{J}_{s} & = & 0,\\
\nabla\cdot\mathcal{J}_{s} & = & 0,\\
-\hbar\epsilon_{\mu\nu\rho\sigma}\nabla^{\rho}\mathcal{J}_{s}^{\sigma} & = & 2s\left[\Pi_{\mu}\mathcal{J}_{s\nu}-\Pi_{\nu}\mathcal{J}_{s\mu}\right].\label{eq:iterative}
\end{eqnarray}
We note that the right- and left-handed Wigner functions have been
totally decoupled with each other.

In this work, we will follow the similar procedure given in Ref.~\citep{Yang:2020mtz, Hidaka:2022dmn}
and make semi-classical expansion to solve the Wigner equation above.
To do that, we need to expand the Wigner functions and the operators
as
\begin{eqnarray}
\mathcal{J}_{s\mu} & = & \sum_{m=0}^{\infty}\hbar^{m}\mathcal{J}_{s\mu}^{(m)},\ \Pi_{\mu}=\sum_{m=0}^{\infty}\hbar^{m}\Pi_{\mu}^{(m)},\ \nabla_{\mu}=\sum_{m=0}^{\infty}\hbar^{m}\nabla_{\mu}^{(m)},
\end{eqnarray}
where the superscript denotes the order of power in the expansion.
From Taylor series expansion of spherical Bessel functions $j_{0}(z)$
and $j_{1}(z)$, we find that the operators $\nabla^{\mu}$ and $\Pi^{\mu}$
includes only even orders. The zeroth order is very simple and given
by
\begin{eqnarray}
\nabla_{\mu}^{(0)} & = & \partial_{\mu}^{x}-F_{\mu\nu}\partial_{p}^{\nu},\nonumber \\
\Pi_{\mu}^{(0)} & = & p_{\mu},
\end{eqnarray}
For the orders of $m>0$, we have the expressions
\begin{eqnarray}
\nabla_{\mu}^{(2m)} & = & \frac{(-1)^{m+1}}{2^{2m}(2m+1)!}\Delta^{2m}F_{\mu\nu}\partial_{p}^{\nu},\nonumber \\
\Pi_{\mu}^{(2m)} & = & \frac{(-1)^{m}m}{2^{2m-1}(2m+1)!}\Delta^{2m-1}F_{\mu\nu}\partial_{p}^{\nu},
\end{eqnarray}
Plugging these expansions into the Wigner equation (\ref{eq:iterative})
and make them hold order by order, we can have
\begin{eqnarray}
\sum_{m=0}^{[n/2]}\Pi^{(2m)}\cdot\mathcal{J}_{s}^{(n-2m)} & = & 0,\\
\sum_{m=0}^{[n/2]}\nabla^{(2m)}\cdot\mathcal{J}_{s}^{(n-2m)} & = & 0,\\
-\sum_{m=0}^{[(n-1)/2]}\epsilon_{\mu\nu\rho\sigma}\nabla^{(2m)\rho}\mathcal{J}_{s}^{(n-2m-1)\sigma} & = & 2s\sum_{m=0}^{[n/2]}\left[\Pi_{\mu}^{(2m)}\mathcal{J}_{s\nu}^{(n-2m)}-\Pi_{\nu}^{(2m)}\mathcal{J}_{s\mu}^{(n-2m)}\right],\label{eq:iterative_expansion}
\end{eqnarray}
where the $[n/2]$ represents the largest integer bounded by $n/2$.
If we define $\sum_{m=0}^{[-1/2]}=0$, Eq. (\ref{eq:iterative_expansion})
also works for $n=0$. Since we restrict ourselves up to the second
order in this work, it will be helpful to give the explicit form of
Wigner equation up to second order. The zeroth-order equations reads
\begin{eqnarray}
p\cdot\mathcal{J}_{s}^{(0)} & = & 0,\label{eq:mass_shell_0}\\
\nabla^{(0)}\cdot\mathcal{J}_{s}^{(0)} & = & 0,\label{eq:evolution_0}\\
p_{\mu}\mathcal{J}_{s\nu}^{(0)}-p_{\nu}\mathcal{J}_{s\mu}^{(0)} & = & 0,\label{eq:iter_0}
\end{eqnarray}
where only zeroth-order Wigner function $\mathcal{J}_{s\mu}^{(0)}$
is involved. The first-order equations are given by
\begin{eqnarray}
p\cdot\mathcal{J}_{s}^{(1)} & = & 0,\label{eq:mass_shell_1}\\
\nabla^{(0)}\cdot\mathcal{J}_{s}^{(1)} & = & 0,\label{eq:evolution_1}\\
p_{\mu}\mathcal{J}_{s\nu}^{(1)}-p_{\nu}\mathcal{J}_{s\mu}^{(1)} & = & -\frac{s}{2}\epsilon_{\mu\nu\rho\sigma}\nabla^{(0)\rho}\mathcal{J}_{s}^{(0)\sigma},\label{eq:iter_1}
\end{eqnarray}
where both the first- and zeroth-order Wigner functions $\mathcal{J}_{s\mu}^{(1)}$
and $\mathcal{J}_{s\mu}^{(0)}$ are involved. The second-order equations
will include new operators associated with the derivative of EM field
\begin{eqnarray}
\nabla_{\mu}^{(2)} & = & \frac{1}{24}\Delta^{2}F_{\mu\nu}\partial_{p}^{\nu},\\
\Pi_{\mu}^{(2)} & = & -\frac{1}{12}\Delta F_{\mu\nu}\partial_{p}^{\nu}.
\end{eqnarray}
and all the Wigner functions with order $n\le2$, i.e., $\mathcal{J}_{s\mu}^{(2)}$,
$\mathcal{J}_{s\mu}^{(1)}$ and $\mathcal{J}_{s\mu}^{(0)}$. The expressions
can be written as
\begin{eqnarray}
p\cdot\mathcal{J}_{s}^{(2)} & = & -\Pi^{(2)}\cdot\mathcal{J}_{s}^{(0)},\label{eq:mass_shell_2}\\
\nabla^{(0)}\cdot\mathcal{J}_{s}^{(2)} & = & -\nabla^{(2)}\cdot\mathcal{J}_{s}^{(0)},\label{eq:evolution_2}\\
p_{\mu}\mathcal{J}_{s\nu}^{(2)}-p_{\nu}\mathcal{J}_{s\mu}^{(2)} & = & -\frac{s}{2}\epsilon_{\mu\nu\rho\sigma}\nabla^{(0)\rho}\mathcal{J}_{s}^{(1)\sigma}-\left[\Pi_{\mu}^{(2)}\mathcal{J}_{s\nu}^{(0)}-\Pi_{\nu}^{(2)}\mathcal{J}_{s\mu}^{(0)}\right].\label{eq:iter_2}
\end{eqnarray}

We note that the zeroth- and first-order Wigner equations are the
same as those under constant EM field used in Ref. \citep{Yang:2020mtz, Hidaka:2022dmn}
while the second-order Wigner equations have included the derivative
terms on EM field which is different from the case under constant
EM field.

For simplicity of notations, we will omit the subscript $s$ in $\mathcal{J}_{s}$
in Sec. \ref{sec:Wigner-function-at} and recover it in Sec. \ref{sec:Vector-and-axial}.

\section{Wigner function at global equilibrium\label{sec:Wigner-function-at}}

In this section, we will solve the Wigner equations to obtain the
Wigner functions at global equilibrium. Let us start from the Eqs.(\ref{eq:iter_0}, \ref{eq:iter_1}, \ref{eq:iter_2})
by multiplying them with $p^{\nu}$ and using Eqs. (\ref{eq:mass_shell_0},\ref{eq:mass_shell_1},\ref{eq:mass_shell_2}),
respectively, we obtain
\begin{eqnarray}
p^{2}\mathcal{J}_{\mu}^{(0)} & = & 0,\\
p^{2}\mathcal{J}_{\mu}^{(1)} & = & \frac{s}{2}\epsilon_{\mu\nu\rho\sigma}p^{\nu}\nabla^{(0)\rho}\mathcal{J}^{(0)\sigma},\\
p^{2}\mathcal{J}_{\mu}^{(2)} & = & \frac{s}{2}\epsilon_{\mu\nu\rho\sigma}p^{\nu}\nabla^{(0)\rho}\mathcal{J}^{(1)\sigma}-p_{\mu}\Pi^{(2)}\cdot\mathcal{J}^{(0)}+p^{\nu}\left[\Pi_{\mu}^{(2)}\mathcal{J}_{\nu}^{(0)}-\Pi_{\nu}^{(2)}\mathcal{J}_{\mu}^{(0)}\right],
\end{eqnarray}
These equations indicate on-shell conditions for chiral Wigner functions
and result in the general expressions
\begin{eqnarray}
\mathcal{J}_{\mu}^{(0)} & = & J_{\mu}^{(0)}\delta(p^{2}),\label{eq:J_0}\\
\mathcal{J}_{\mu}^{(1)} & = & J_{\mu}^{(1)}\delta(p^{2})+\frac{s}{2p^{2}}\epsilon_{\mu\nu\rho\sigma}p^{\nu}\nabla^{(0)\rho}\mathcal{J}^{(0)\sigma},\label{eq:J_1}\\
\mathcal{J}_{\mu}^{(2)} & = & J_{\mu}^{(2)}\delta(p^{2})+\frac{s}{2p^{2}}\epsilon_{\mu\nu\rho\sigma}p^{\nu}\nabla^{(0)\rho}\mathcal{J}^{(1)\sigma}\nonumber \\
 &  & -\frac{p_{\mu}}{p^{2}}\Pi^{(2)}\cdot\mathcal{J}^{(0)}+\frac{p^{\nu}}{p^{2}}\left[\Pi_{\mu}^{(2)}\mathcal{J}_{\nu}^{(0)}-\Pi_{\nu}^{(2)}\mathcal{J}_{\mu}^{(0)}\right],\label{eq:J_2}
\end{eqnarray}
where $J_{\mu}^{(n)}$ are unknown functions of $x,p$ with non-singularity
at $p^{2}=0$. These unknown functions can be further constrained
by Wigner equations. Substituting the expressions (\ref{eq:J_0}) -(\ref{eq:J_2})
into Eqs.(\ref{eq:mass_shell_0}), (\ref{eq:mass_shell_1}), and (\ref{eq:mass_shell_2}),
respectively, leads to constraint equations
\begin{eqnarray}
p^{\mu}J_{\mu}^{(n)}\delta(p^{2}) & = & 0.\ \ \ \ \ (n=0,1,2)\label{eq:constrain_0}
\end{eqnarray}
To determine $J_{\mu}^{(n)}$ further, it is convenient to decompose
$J_{\mu}^{(n)}$ into two parts
\begin{eqnarray}
J_{\mu}^{(n)} & = & p_{\mu}f^{(n)}+X_{\mu}^{(n)}.\label{p-X}
\end{eqnarray}
where the first term satisfies the constraint equation (\ref{eq:constrain_0})
automatically due to on-shell Dirac delta function and the second
term satisfies $p^{\mu}X_{\mu}^{(n)}=0$ even without the on-shell
Dirac delta function.

\subsection{Zeroth and first order solution}
For the zeroth-order result, substituting the Eq. (\ref{eq:J_0})
together with (\ref{p-X}) into Eq. (\ref{eq:iter_0}), we find $X_{\mu}^{(0)}$
vanishs and the zeroth-order Wigner function is given by
\begin{eqnarray}
\mathcal{J}_{\mu}^{(0)} & = & p_{\mu}f^{(0)}\delta(p^{2}).\label{eq:J_0_result}
\end{eqnarray}
Since we intend to find the solution in global equilibrium with vorticity
and EM field in what follows, it is very natural to choose the zeroth-order
distribution function $f^{(0)}$ as the Fermi-Dirac distribution for
free particles
\begin{eqnarray}
f^{(0)} & = & \left\{ \begin{array}{ll}
\frac{1}{4\pi^{3}}\frac{1}{e^{\beta\cdot p-\bar{\mu}_{s}}+1},\hspace{1.8cm}(p_{0}>0)\\
\ \\
\frac{1}{4\pi^{3}}\left(\frac{1}{e^{-\beta\cdot p+\bar{\mu}_{s}}+1}-1\right),\hspace{0.5cm}(p_{0}<0).
\end{array}\right.\label{eq:f-0-free}
\end{eqnarray}
where $\beta^\mu$ is the four-temperature vector in connection with the
fluid velocity by $\beta^{\mu}={u^{\mu}}/{T}$ and $\bar{\mu}_{s}={\mu_{s}}/{T}$
denotes the right- or left-handed chemical potential $\mu_{s}$ scaled
by temperature $T$. {The factor $1/(4\pi^3)$ is determined by mapping $\int d^4p (u\cdot p) f^{(0)} $ to number density of non-interacting fermionic gas. The additional $-1$ in $f^{(0)}$ when $p_{0}<0$ comes from the vacuum contribution \cite{Gao:2019zhk}. }
The vector and axial chemical potential can be
related to the right- and left-handed chemical potential by
\begin{equation}
    \bar{\mu}_{s}=\bar{\mu}+s\bar{\mu}_{5}.
\end{equation}
Inserting Eq. (\ref{eq:J_0_result}) into Eq. (\ref{eq:evolution_0})
gives rise to the transport equation of $f^{(0)}$,
\begin{eqnarray}
p^{\mu}\nabla_{\mu}^{(0)}f^{(0)}\delta(p^{2}) & = & 0,
\end{eqnarray}
Given the specific expression of $f^{(0)}$ in Eq.~(\ref{eq:f-0-free}),
this equation hold only under the constraint condition \citep{Gao:2012ix,Yang:2020mtz}
\begin{eqnarray}
\partial_{\mu}\beta_{\nu}+\partial_{\nu}\beta_{\mu} & = & 0,\label{eq:constrain_1}\\
\partial_{\mu}\bar{\mu}+F_{\mu\lambda}\beta^{\lambda} & = & 0,\label{eq:constrain_2}\\
\partial_{\mu}\bar{\mu}_{5} & = & 0.\label{eq:constrain_3}
\end{eqnarray}
The Eq.~(\ref{eq:constrain_3}) means the axial chemical potential
must be a constant. From the equation (\ref{eq:constrain_1}), we
can verify that the thermal vorticity
\begin{eqnarray}
\Omega_{\mu\nu} & = & \frac{1}{2}(\partial_{\mu}\beta_{\nu}-\partial_{\nu}\beta_{\mu}).
\end{eqnarray}
is a constant tensor, i.e., $\partial_{\rho}\Omega_{\mu\nu}=0$, {at the global equilibrium}. For
the equation (\ref{eq:constrain_2}), we act partial derivative $\partial_{\nu}$
on both sides and obtain
\begin{eqnarray}
\partial_{\mu}\partial_{\nu}\bar{\mu}+\partial_{\mu}(F_{\nu\lambda}\beta^{\lambda}) & = & 0.
\end{eqnarray}
Commutativity of partial derivatives lead to the integrability condition
\begin{eqnarray}
F_{\mu\lambda}\Omega_{\nu}^{\;\lambda}-F_{\nu\lambda}\Omega_{\mu}^{\;\lambda} & = & -\beta^{\lambda}(\partial_{\lambda}F_{\mu\nu}),\label{eq:integrability_1}
\end{eqnarray}
where we have used the Bianchi identity
\begin{eqnarray}
\partial_{\mu}F_{\nu\rho}+\partial_{\nu}F_{\rho\mu}+\partial_{\rho}F_{\mu\nu} & = & 0.\label{eq:Maxwell_equation}
\end{eqnarray}
The derivative terms on the right-handed side of the integrability
condition (\ref{eq:integrability_1}) is the main difference from
our previous work \citep{Yang:2020mtz}, in which this term vanishes
for the constant EM field.

Given the zeroth-order result (\ref{eq:J_0_result}), the first-order
chiral Wigner function can be obtained from Eq. (\ref{eq:J_1}),
\begin{eqnarray}
\mathcal{J}_{\mu}^{(1)} & = & p_{\mu}f^{(1)}\delta(p^{2})+X_{\mu}^{(1)}\delta(p^{2})+s\tilde{F}_{\mu\nu}p^{\nu}f^{(0)}\delta^{\prime}(p^{2}).\label{eq:J_1perp_X}
\end{eqnarray}
where $\delta^{\prime}(p^{2})=\partial\delta(p^{2})/\partial p^{2}$
and
\begin{equation}
    \tilde{F}_{\mu\nu}=\frac{1}{2}\epsilon_{\mu\nu\rho\sigma}F^{\rho\sigma}.
\end{equation}
Inserting this first-order Wigner funciton (\ref{eq:J_1perp_X}) into
the Eq.~(\ref{eq:iter_1}), we find that the term $X_{\mu}^{(1)}$
must take the form of 
\begin{eqnarray}
X_{\mu}^{(1)} & = & -\frac{s}{2}\tilde{\Omega}_{\mu\lambda}p^{\lambda}f^{(0)\prime},\label{X-mu-1}
\end{eqnarray}
where $f^{(0)\prime}=\partial f^{(0)}/\partial(\beta\cdot p)$ and
\begin{equation}
    \tilde{\Omega}_{\mu\nu}=\frac{1}{2}\epsilon_{\mu\nu\rho\sigma}\Omega^{\rho\sigma}.
\end{equation}
Now we substitute the expression (\ref{eq:J_1perp_X}) with the result
(\ref{X-mu-1}) into the first-order transport equation(\ref{eq:evolution_1}),
we find that the second and third terms in Eq.~(\ref{eq:J_1perp_X}) cancel
each other and only the first term remains
\begin{eqnarray}
\nabla^{(0)}\cdot[pf^{(1)}\delta(p^{2})] & = & 0,\label{eq:evolution_1_part_1}
\end{eqnarray}
It turns out that $f^{(1)}$ must vanish through detailed derivation
which will be given in Sec.\ref{sec:general-solution}. Thus we obtain
the final first-order Wigner function
\begin{eqnarray}
\mathcal{J}_{\mu}^{(1)} & = & -\frac{s}{2}\tilde{\Omega}_{\mu\lambda}p^{\lambda}f^{(0)\prime}\delta(p^{2})+s\tilde{F}_{\mu\nu}p^{\nu}f^{(0)}\delta^{\prime}(p^{2}).\label{eq:J1_result}
\end{eqnarray}

\subsection{Second order solution}
The second-order Wigner function follows in similar procedure. Substituting
the zeroth- and first-order Wigner functions (\ref{eq:J_0_result})
and (\ref{eq:J_1perp_X}) into the Eq. (\ref{eq:J_2}) gives rise
to the second-order Wigner function
\begin{eqnarray}
\mathcal{J}_{\mu}^{(2)} & = & p_{\mu}f^{(2)}\delta(p^{2})+X_{\mu}^{(2)}\delta(p^{2})+\tilde{\mathcal{J}}_{\mu}^{(2)}+\Delta{\mathcal{J}}_{\mu}^{(2)}\label{eq:J_2_result},
\end{eqnarray}
where $\tilde{\mathcal{J}}_{\mu}^{(2)}$ and $\Delta{\mathcal{J}}_{\mu}^{(2)}$
are derived from the last three terms in Eq.~(\ref{eq:J_2}). The contribution
$\tilde{\mathcal{J}}_{\mu}^{(2)}$ is the result for the constant
EM field and has already been obtain in Refs.~\citep{Yang:2020mtz,Hidaka:2022dmn}
\begin{eqnarray}
\tilde{\mathcal{J}}_{\mu}^{(2)} & = & -\frac{1}{4}p_{\mu}\Omega_{\nu\rho}p^{\nu}\Omega^{\lambda\rho}p_{\lambda}f^{(0)\prime\prime}\delta^{\prime}(p^{2})
-\frac{1}{4}\Omega_{\mu\nu}\Omega^{\lambda\nu}p_{\lambda}f^{(0)\prime\prime}\delta(p^{2})\nonumber \\
 &  & +\frac{1}{2}p_{\mu}F_{\nu\gamma}p^{\nu}\Omega^{\lambda\gamma}p_{\lambda}f^{(0)\prime}\delta^{\prime\prime}(p^{2})
 +\frac{1}{2}\left(F_{\mu\gamma}\Omega^{\lambda\gamma}+F^{\lambda\gamma}\Omega_{\mu\gamma}\right)p_{\lambda}f^{(0)\prime}\delta^{\prime}(p^{2})\nonumber \\
 &  & -\frac{1}{3}p_{\mu}F_{\nu\rho}p^{\nu}F^{\lambda\rho}p_{\lambda}f^{(0)}\delta^{\prime\prime\prime}(p^{2})
 -F_{\mu\nu}F^{\lambda\nu}p_{\lambda}f^{(0)}\delta^{\prime\prime}(p^{2})\label{tJ-2}.
\end{eqnarray}
While $\Delta{\mathcal{J}}_{\mu}^{(2)}$ involves only derivative
terms of EM field,
\begin{eqnarray}
\Delta{\mathcal{J}}_{\mu}^{(2)} & = & -\frac{2}{3}(\partial^{\lambda}F_{\mu\lambda})f^{(0)}\delta^{\prime}(p^{2})
-\frac{1}{12}\beta_{\lambda}(\partial^{\lambda}F_{\mu\nu})\beta^{\nu}f^{(0)\prime\prime}\delta(p^{2})\nonumber \\
 &  & -\frac{1}{12}[p_{\mu}(\partial^{\nu}F_{\nu\lambda})+2p^{\nu}\partial_{\mu}F_{\nu\lambda}{+4p^{\nu}\partial_{\lambda}F_{\mu\nu}}]\beta^{\lambda}f^{(0)\prime}\delta^{\prime}(p^{2})\nonumber \\
 &  & +\frac{1}{3}[p_{\mu}(\partial^{\lambda}F_{\lambda\nu})p^{\nu}-p_{\lambda}(\partial^{\lambda}F_{\mu\nu})p^{\nu}]f^{(0)}\delta^{\prime\prime}(p^{2})\nonumber \\
 &  & -\frac{1}{6}p_{\mu}\beta^{\lambda}(\partial_{\lambda}F_{\nu\rho})p^{\nu}\beta^{\rho}f^{(0)\prime\prime}\delta^{\prime}(p^{2})
 -\frac{1}{6}p_{\mu}p^{\lambda}(\partial_{\lambda}F_{\nu\rho})p^{\nu}\beta^{\rho}f^{(0)\prime}\delta^{\prime\prime}(p^{2}).\label{DJ-2}
\end{eqnarray}
It should be
noted that we have used the integrability condition (\ref{eq:integrability_1})
to obtain the expressions (\ref{tJ-2}) and (\ref{DJ-2}).
The Eq.~(\ref{DJ-2}) is the main research object in our present work.

The term $X_{\mu}^{(2)}$ can be determined by Eq.~(\ref{eq:iter_2})
with the expressions in Eqs.~(\ref{eq:J_0_result}, \ref{eq:J_1perp_X}).
We find that it must vanish.

\subsection{Discussion on $f^{(2)}$ in Eq.~(\ref{eq:J_2_result})}
With vanishing $X_{\mu}^{(2)}$, plugging
Eq. (\ref{eq:J_2_result}) into the second-order transport equation
(\ref{eq:evolution_2}), we have
\begin{eqnarray}
\nabla^{(0)}\cdot[pf^{(2)}\delta(p^{2})]
-\frac{1}{12}\beta^{\lambda}\Omega^{\mu\rho}p_{\rho}\beta_{\kappa}(\partial^{\kappa}F_{\mu\lambda})f^{(0)\prime\prime\prime}\delta\left(p^{2}\right)
+\frac{1}{24}\beta_{\kappa}\beta_{\lambda}p^{\mu}\beta^{\nu}(\partial^{\kappa}\partial^{\lambda}F_{\mu\nu})f^{(0)\prime\prime\prime}\delta\left(p^{2}\right)=0,\label{qke-2}
\end{eqnarray}
In Ref.~\citep{Mameda:2023ueq}, the author choose the specific solution
with $f^{(2)}=0$ to make the first term vanish and impose extra constraint
conditions $\beta\cdot\partial F_{\mu\nu}=0$ and $\beta\cdot\partial\beta_{\mu}=0$
to make the last two terms vanish. With these specific solution and
extra constraints, the equation is satisfied automatically. In this
work, we will release these extra constraints and manage to find more
general solutions.

The key point that we are able to find a general solution is that
we can cast the Eq.~(\ref{qke-2}) into the form
\begin{eqnarray}
 &  & \nabla^{(0)}\cdot[pf^{(2)}\delta(p^{2})]-\nabla^{(0)}\cdot[p\Delta f^{(2)}\delta(p^{2})]=0, \label{qke-2-a}
\end{eqnarray}
where the second term above comes from the last two terms of Eq. (\ref{qke-2})
and $\Delta f^{(2)}$ is given by
\begin{eqnarray}
\Delta f^{(2)} & = & -\frac{1}{24}F_{\rho\lambda}\Omega_{\sigma}^{\ \lambda}\beta^{\rho}\beta^{\sigma}f^{(0)\prime\prime\prime}
-\frac{1}{12}\text{\ensuremath{\Omega}}_{\rho\lambda}\Omega_{\sigma}^{\ \lambda}p^{\rho}\beta^{\sigma}f^{(0)\prime\prime\prime},\label{eq:Delta_f2}
\end{eqnarray}
Without extra constraints, the second term in Eq. (\ref{qke-2-a})
does not vanish and we cannot set $f^{(2)}=0$. However, because the
two terms share the same structure as shown in Eq. (\ref{qke-2-a}),
we can divide
\begin{equation}
    f^{(2)}=\tilde{f}^{(2)}+\Delta f^{(2)},
\end{equation}
to cancel
the second term in Eq. (\ref{qke-2-a}) and obtain the transport equation
for $\tilde{f}^{(2)}$
\begin{eqnarray}
 &  & \nabla^{(0)}\cdot[p\tilde{f}^{(2)}\delta(p^{2})]=0\label{eq:evolution_2_part_2}
\end{eqnarray}
{Obvisouly, $\tilde{f}^{(2)}=0$ is one specific solution for the above equations.}
If we simply choose $\tilde{f}^{(2)}=0$, then the
second-order chiral Wigner equation is given by
\begin{eqnarray}
\mathcal{J}_{\mu}^{(2)} & = & p_{\mu}\Delta f^{(2)}\delta(p^{2})+\tilde{\mathcal{J}}_{\mu}^{(2)}+\Delta{\mathcal{J}}_{\mu}^{(2)}.\label{eq:J_2_result-tf0}
\end{eqnarray}
{In Sec.~\ref{sec:Vector-and-axial}, \ref{sec:energy-momentum-tensor} and \ref{sec:polarization}, we compute the currents and energy-momentum tensor by simply setting $\tilde{f}^{(2)}=0$ for simplicity.}
In Sec.~\ref{sec:general-solution}, we will verify that $\tilde{f}^{(2)}$ could have general non-vanishing solution { and discuss the possible corrections from $\tilde{f^{(2)}}$.}

\subsection{Tensor decomposition along the $u^\mu$}
Given the chiral Wigner functions, we can get the currents and energy-momentum
tensor directly by integrating them with corresponding moments. We
will carry out these integration in the next two sections. At the
end of this section, it is a good place to introduce the decomposition
of vectors and tensors along the direction of fluid velocity $u^{\mu}$
with normalization $u^{2}=1$. With time-like vector $u^{\mu}$, we
can define the projector orthogonal to $u^{\mu}$
\begin{eqnarray}
\Delta^{\mu\nu} & = & g^{\mu\nu}-u^{\mu}u^{\nu}
\end{eqnarray}
with which we can decompose any vector, taking the momentum $p^{\mu}$
as an example, as
\begin{eqnarray}
p^{\mu}=(u\cdot p)u^{\mu}+\Delta^{\mu\nu}p_{\nu}
\end{eqnarray}
The antisymmetric tensors $\Omega^{\mu\nu}$ and $F^{\mu\nu}$ can
be decomposition into
\begin{eqnarray}
T\Omega^{\mu\nu} & = & \varepsilon^{\mu}u^{\nu}-\varepsilon^{\nu}u^{\mu}+\epsilon^{\mu\nu\rho\sigma}u_{\rho}\omega_{\sigma},\\
F^{\mu\nu} & = & E^{\mu}u^{\nu}-E^{\nu}u^{\mu}+\epsilon^{\mu\nu\rho\sigma}u_{\rho}B_{\sigma}.\label{eq:decompose_F}
\end{eqnarray}
with $\varepsilon^{\mu}=T\Omega^{\mu\nu}u_{\nu}$, $\omega^{\mu}=T\epsilon^{\mu\nu\rho\sigma}u_{\nu}\Omega_{\rho\sigma}/2$,
$E^{\mu}=F^{\mu\nu}u_{\nu}$, $B^{\mu}=\epsilon^{\mu\nu\rho\sigma}u_{\nu}F_{\rho\sigma}/2$.
Using the decompositions of $\Omega^{\mu\nu}$ and $F^{\mu\nu}$ above,
we can write the left-hand side of integrability condition (\ref{eq:integrability_1})
in terms of $\varepsilon^{\mu}$, $\omega^{\mu}$, $E^{\mu}$, and
$B^{\mu}$, i.e.,
\begin{eqnarray}
E_{\mu}\varepsilon_{\nu}-E_{\nu}\varepsilon_{\mu}+\omega_{\mu}B_{\nu}-\omega_{\nu}B_{\mu}+\left(u_{\mu}\epsilon_{\nu\rho\sigma\lambda}-u_{\nu}\epsilon_{\mu\rho\sigma\lambda}\right)u^{\rho}\left(E^{\sigma}\omega^{\lambda}-\varepsilon^{\sigma}B^{\lambda}\right) & = & -u^{\lambda}\partial_{\lambda}F_{\mu\nu}
\end{eqnarray}
where we multiply both sides with $T$. Contracting both sides with
$u^{\nu}$ and $\epsilon^{\alpha\beta\mu\nu}u_{\beta}$ lead to, respectively,
\begin{eqnarray}
\epsilon_{\mu\nu\rho\sigma}u^{\nu}\left(E^{\rho}\omega^{\sigma}-\varepsilon^{\rho}B^{\sigma}\right) & = & u^{\nu}u^{\lambda}\partial_{\lambda}F_{\mu\nu},\label{dF-E-omega}\\
\epsilon_{\mu\nu\rho\sigma}u^{\nu}\left(E^{\rho}\varepsilon^{\sigma}+\omega^{\rho}B^{\sigma}\right) & = & -u^{\nu}u^{\lambda}\partial_{\lambda}\tilde{F}_{\mu\nu}\label{dtF-B-omega}
\end{eqnarray}
Sometimes, we need write both sides of integrability condition (\ref{eq:integrability_1})
in terms of $\varepsilon^{\mu}$, $\omega^{\mu}$, $E^{\mu}$, and
$B^{\mu}$, i.e.,
\begin{eqnarray}
-\epsilon_{\mu\nu\alpha\gamma}E^{\alpha}\omega^{\gamma}+\omega_{\mu}B_{\nu}-\omega_{\nu}B_{\mu}=-u_{\nu}u^{\lambda}\partial_{\lambda}E_{\mu}+u_{\mu}u^{\lambda}\partial_{\lambda}E_{\nu}-\epsilon_{\mu\nu\rho\sigma}u^{\rho}u^{\lambda}\partial_{\lambda}B^{\sigma}
\end{eqnarray}
Contracting both sides with $u^{\nu}$ and $\epsilon^{\alpha\beta\mu\nu}u_{\beta}$
lead to, respectively,
\begin{eqnarray}
\varepsilon\cdot Eu_{\mu}+\epsilon_{\mu\nu\rho\sigma}u^{\nu}E^{\rho}\omega^{\sigma} & = & u^{\lambda}\partial_{\lambda}E_{\mu},\label{eq:partial_t_E}\\
\varepsilon\cdot Bu_{\mu}+\epsilon_{\mu\nu\rho\sigma}u^{\nu}B^{\rho}\omega^{\sigma} & = & u^{\lambda}\partial_{\lambda}B_{\mu}.\label{eq:partial_t_B}
\end{eqnarray}
Note that the three terms $u_{\lambda}\partial^{\lambda}E^{\mu}$,
$\varepsilon\cdot Eu^{\mu}$, and $\epsilon^{\mu\nu\rho\sigma}u_{\nu}E_{\rho}\omega_{\sigma}$
are linearly related and only two of them are independent. It is the
same for $u_{\lambda}\partial^{\lambda}B^{\mu}$, $\varepsilon\cdot Bu^{\mu}$,
and $\epsilon^{\mu\nu\rho\sigma}u_{\nu}B_{\rho}\omega_{\sigma}$.
These points will be important to consider the general solution in
Sec. \ref{sec:general-solution}.

\section{Vector and axial current\label{sec:Vector-and-axial}}

The currents with chirality $s$ can be obtained by integrating the
chiral Wigner function over momentum.
\begin{eqnarray}
j_{s}^{\mu} & = & \int d^{4}p\mathcal{J}_{s}^{\mu}.\label{eq:j_integrate}
\end{eqnarray}
Performing the similar procedure and method used in Refs.~\citep{Yang:2020mtz,Hidaka:2022dmn},
we obtain the $j_{s}^{\mu}$ up to second order
\begin{eqnarray}
j_{s}^{(0)\mu} & = & \frac{\mu_{s}}{6\pi^{2}}(\pi^{2}T^{2}+\mu_{s}^{2})u^{\mu},\\
j_{s}^{(1)\mu} & = & \frac{s}{12\pi^{2}}(\pi^{2}T^{2}+3\mu_{s}^{2})\omega^{\mu}+\frac{s\mu_{s}}{4\pi^{2}}B^{\mu},\\
j_{s}^{(2)\mu} & = & -\frac{\mu_{s}}{4\pi^{2}}(\varepsilon^{2}+\omega^{2})u^{\mu}-\frac{1}{8\pi^{2}}(\varepsilon\cdot E+\omega\cdot B)u^{\mu}-\frac{C_{s}}{24\pi^{2}}(E^{2}+B^{2})u^{\mu}\nonumber \\
 &  & {-\frac{1}{16\pi^{2}}\epsilon^{\mu\nu\rho\sigma}u_{\nu}(E_{\rho}\omega_{\sigma}+\varepsilon_{\rho}B_{\sigma})}-\frac{C_{s}}{12\pi^{2}}\epsilon^{\mu\nu\rho\sigma}u_{\nu}E_{\rho}B_{\sigma}+\Delta^{\textrm{\tiny{I}}}j_{s}^{(2)\mu}+\Delta^{\textrm{\tiny{II}}}j_{s}^{(2)\mu}
\end{eqnarray}
The detail of the integration is given in Appendix~\ref{sec:integral-for-currents}.
The zeroth- and first-order results $j_{s}^{(0)\mu}$ and $j_{s}^{(1)\mu}$
are both well-known. In contrast with the result given in Ref.~\citep{Yang:2020mtz},
the second-order current $j_{s}^{(2)\mu}$ receive extra contributions
$\Delta^{\textrm{\tiny{I}}}j_{s}^{(2)\mu}$ and $\Delta^{\textrm{\tiny{II}}}j_{s}^{(2)\mu}$.
The term $\Delta^{\textrm{\tiny{I}}}j_{s}^{(2)\mu}$ comes from the
contribution associated with $\Delta f^{(2)}$ in Eq.(\ref{eq:J_2_result-tf0})
and is given by
\begin{eqnarray}
\Delta^{\textrm{\tiny{I}}}j_{s}^{(2)\mu} & = & \frac{\mu_{s}}{4\pi^{2}}u^{\mu}\varepsilon^{2}+\frac{1}{24\pi^{2}}u^{\mu}\varepsilon\cdot E-\frac{\mu_{s}}{12\pi^{2}}\epsilon^{\mu\nu\rho\sigma}u_{\nu}\varepsilon_{\rho}\omega_{\sigma}
\end{eqnarray}
The term $\Delta^{\textrm{\tiny{II}}}j_{s}^{(2)\mu}$ comes from the
contribution $\Delta{\mathcal{J}}_{\mu}^{(2)}$ and involves only
the derivative terms with EM field
\begin{eqnarray}
\Delta^{\textrm{\tiny{II}}}j_{s}^{(2)\mu} & = & -\frac{1}{12}\left(\kappa_{s}^{\epsilon}+\frac{1}{\pi^{2}}\right)\partial_{\lambda}F^{\mu\lambda}+\frac{1}{24\pi^{2}}u_{\nu}u^{\mu}\partial_{\lambda}F^{\lambda\nu}{+\frac{1}{48\pi^{2}}u_{\nu}u^{\lambda}\partial_{\lambda}F^{\mu\nu}},\label{eq:j_2_mu}
\end{eqnarray}
The coefficients $C_{s}$ and $\kappa_{s}^{\epsilon}$ are defined
by, respectively,
\begin{eqnarray}
C_{s} & = & \frac{1}{T}\int_{0}^{\infty}\frac{dy}{y}\left[\frac{e^{y-\bar{\mu}_{s}}}{(e^{y-\bar{\mu}_{s}}+1)^{2}}-\frac{e^{y+\bar{\mu}_{s}}}{(e^{y+\bar{\mu}_{s}}+1)^{2}}\right],\\
\kappa_{s}^{\epsilon} & = & \frac{4\pi^{\frac{3-\epsilon}{2}}T^{-\epsilon}}{\Gamma\left(\frac{3-\epsilon}{2}\right)(2\pi)^{3-\epsilon}}\int dyy^{-1-\epsilon}\left(\frac{1}{e^{y-\bar{\mu}_{s}}+1}+\frac{1}{e^{y+\bar{\mu}_{s}}+1}-1\right).
\end{eqnarray}
The small parameter $\epsilon=4-d$ appears due to the dimensional
regularization when we calculate the integration in $d$-dimentional
momentum space. The detailed manipulation can be found in Ref.~\citep{Yang:2020mtz}.
Here we just present the final result of $k_{s}^{\epsilon}$ expanded
around $\epsilon=0$,
\begin{eqnarray}
\kappa_{s}^{\epsilon} & = & -\frac{1}{\pi^{2}}\left[\frac{1}{\epsilon}+\ln2+\frac{1}{2}\ln\pi+\psi\left(\frac{3}{2}\right)-\ln T+\hat{\kappa}_{s}\right],
\end{eqnarray}
where $\psi(x)$ is digamma function and $\hat{\kappa}^{\epsilon}$
is given by
\begin{eqnarray}
\hat{\kappa}_{s} & = & \int_{0}^{\infty}dy\ln y\frac{d}{dy}\left[\frac{1}{e^{y-\bar{\mu}_{s}}+1}+\frac{1}{e^{y+\bar{\mu}_{s}}+1}\right].
\end{eqnarray}

We note that the divergent contribution arises from the coefficient
$\kappa_{s}^{\epsilon}$ while it is absent in the constant EM field.
In Ref.~\citep{Mameda:2023ueq}, the author also calculated the derivative
terms on the EM field and obtain the divergent counterpart of the
first term in Eq. (\ref{eq:j_2_mu_v}) in different regularization
scheme. However, we obtain additional contribution, the second and
third terms in Eq. (\ref{eq:j_2_mu_v}), which is the main difference
from Ref.~\citep{Mameda:2023ueq}.

The sum and difference give the vector and axial currents, respectively,
\begin{eqnarray}
j^{\mu} & = & j_{+1}^{\mu}+j_{-1}^{\mu},\ j_{5}^{\mu}=j_{+1}-j_{-1}.\label{eq:j_j5}
\end{eqnarray}
In the following, we will not present the zeroth- and first-order
results any more since they are already well-known. We will only show
the second-order results which includes new contributions compared
with Ref.~\citep{Yang:2020mtz} and Ref.~\citep{Mameda:2023ueq}.

The vector and axial currents at second order are given by
\begin{eqnarray}
j^{(2)\mu} & = & -\frac{\mu}{2\pi^{2}}(\varepsilon^{2}+\omega^{2})u^{\mu}-\frac{1}{4\pi^{2}}(\varepsilon\cdot E+\omega\cdot B)u^{\mu}-\frac{C}{12\pi^{2}}(E^{2}+B^{2})u^{\mu}\nonumber \\
 &  & {-\frac{1}{8\pi^{2}}\epsilon^{\mu\nu\rho\sigma}u_{\nu}(E_{\rho}\omega_{\sigma}+\varepsilon_{\rho}B_{\sigma})}-\frac{C}{6\pi^{2}}\epsilon^{\mu\nu\rho\sigma}u_{\nu}E_{\rho}B_{\sigma}+\Delta^{\textrm{\tiny{I}}}j^{(2)\mu}+\Delta^{\textrm{\tiny{II}}}j^{(2)\mu},\label{eq:j2_c}\\
j_{5}^{(2)\mu} & = & -\frac{\mu_{5}}{2\pi^{2}}(\varepsilon^{2}+\omega^{2})u^{\mu}-\frac{C_{5}}{12\pi^{2}}(E^{2}+B^{2})u^{\mu}-\frac{C_{5}}{6\pi^{2}}\epsilon^{\mu\nu\rho\sigma}u_{\nu}E_{\rho}B_{\sigma}+\Delta^{\textrm{\tiny{I}}}j_{5}^{(2)\mu}+\Delta^{\textrm{\tiny{II}}}j_{5}^{(2)\mu},\label{eq:j2_5c}
\end{eqnarray}
where the terms $\Delta^{\textrm{\tiny{I}}}j^{(2)\mu}$, $\Delta^{\textrm{\tiny{I}}}j_{5}^{(2)\mu}$,
$\Delta^{\textrm{\tiny{II}}}j^{(2)\mu}$ and $\Delta^{\textrm{\tiny{II}}}j_{5}^{(2)\mu}$
are defined as
\begin{eqnarray}
\Delta^{\textrm{\tiny{I}}}j^{(2)\mu} & = & \Delta^{\textrm{\tiny{I}}}j_{+}^{(2)\mu}+\Delta^{\textrm{\tiny{I}}}j_{-}^{(2)\mu}=\frac{\mu}{2\pi^{2}}u^{\mu}\varepsilon^{2}+\frac{1}{12\pi^{2}}u^{\mu}\varepsilon\cdot E-\frac{\mu}{6\pi^{2}}\epsilon^{\mu\nu\rho\sigma}u_{\nu}\varepsilon_{\rho}\omega_{\sigma},\\
\Delta^{\textrm{\tiny{I}}}j_{5}^{(2)\mu} & = & \Delta^{\textrm{\tiny{I}}}j_{+}^{(2)\mu}-\Delta^{\textrm{\tiny{I}}}j_{-}^{(2)\mu}=\frac{\mu_{5}}{2\pi^{2}}u^{\mu}\varepsilon^{2}-\frac{\mu_{5}}{6\pi^{2}}\epsilon^{\mu\nu\rho\sigma}u_{\nu}\varepsilon_{\rho}\omega_{\sigma},\\
\Delta^{\textrm{\tiny{II}}}j^{(2)\mu} & = & \Delta^{\textrm{\tiny{II}}}j_{+}^{(2)\mu}+\Delta^{\textrm{\tiny{II}}}j_{-}^{(2)\mu}=-\frac{1}{6}\left(\kappa^{\epsilon}+\frac{1}{\pi^{2}}\right)\partial_{\lambda}F^{\mu\lambda}+\frac{1}{12\pi^{2}}u_{\nu}u^{\mu}\partial_{\lambda}F^{\lambda\nu}{+\frac{1}{24\pi^{2}}u_{\nu}u^{\lambda}\partial_{\lambda}F^{\mu\nu}},\label{eq:j_2_mu_v}\\
\Delta^{\textrm{\tiny{II}}}j_{5}^{(2)\mu} & = & \Delta^{\textrm{\tiny{II}}}j_{+}^{(2)\mu}-\Delta^{\textrm{\tiny{II}}}j_{-}^{(2)\mu}=-\frac{1}{12}\kappa_{5}\partial_{\lambda}F^{\mu\lambda}\label{eq:j_2_mu_5}
\end{eqnarray}
and the coefficients $C$, $C_{5}$, $\kappa^{\epsilon}$ and $\kappa_{5}$
are given by
\begin{eqnarray}
C & = & \frac{1}{2}(C_{+}+C_{-}),\ C_{5}=\frac{1}{2}(C_{+}-C_{-}),\\
\kappa^{\epsilon} & = & \frac{1}{2}(\kappa_{+}^{\epsilon}+\kappa_{-}^{\epsilon}),\ \ \kappa_{5}=\frac{1}{2}(\kappa_{+}^{\epsilon}-\kappa_{-}^{\epsilon}).
\end{eqnarray}

We note that the second-order vector and axial currents both receive
extra contributions from the varying EM field in space or time.
{Let us compare our current results with those derived
in Ref.~\cite{Mameda:2023ueq}}.
{The $\Delta^{\textrm{\tiny{I}}}j^{(2)\mu}$ and $\Delta^{\textrm{\tiny{I}}}j^{(2)\mu}_5$
originate from the
contribution  $\Delta f^{(2)}$ which is necessary to satisfy
the Wigner equation for the time- and space- dependent EM field. Thus, these terms are absent in cases with constant EM fields \citep{Mameda:2023ueq}.
The $\Delta^{\textrm{\tiny{II}}}j^{(2)\mu}$ and $\Delta^{\textrm{\tiny{II}}}j^{(2)\mu}_5$
originate from the pure derivative terms $\Delta{\mathcal{J}}_{\mu}^{(2)}$,
among which the terms relevant to $\partial_{\lambda}F^{\mu\lambda}$
have already been obtained in Ref.~\citep{Mameda:2023ueq}. Interestingly, the term proportional to
$u^{\lambda}\partial_{\lambda}F^{\mu\nu}$ in $\Delta^{\textrm{\tiny{II}}}j^{(2)\mu}$ is new in the comparison with the results in Ref.~\citep{Mameda:2023ueq} due to the extra constraint conditions.}
We notice that the $u_{\nu}u^{\lambda}\partial_{\lambda}F^{\mu\nu}$
can be written as $\epsilon_{\mu\nu\rho\sigma}u^{\nu}(E^{\rho}\omega^{\sigma}-\varepsilon^{\rho}B^{\sigma})$ by the integrability condition (\ref{dF-E-omega}).
According to the Maxwell's equation $\partial_{\lambda}F^{\mu\lambda}=j^{\mu}_b$,
we can identify the terms proportional to $\partial_{\lambda}F^{\mu\lambda}$ in $\Delta^{\textrm{\tiny{II}}}j^{(2)\mu}$ as the
induced currents directly from the background currents $j^{\mu}_b$ which generate
the background EM field.
 We also note that
the derivative terms have brought divergence coefficients $\kappa^{\epsilon}$
which is absent for constant EM field. These divergence comes from
the vacuum contribution of the distribution function in Eq.~(\ref{eq:f-0-free})
and can be removed by electric charge renormalization. It is interesting
that the divergence only appear for vector current while the coefficient
of axial current $\kappa_{5}$ is finite and receives on divergence.
Since the infinity has been cancelled in the definition of $\kappa_{5}$,
we have dropped the superscript $\epsilon$ in $\kappa_{5}$.

{For the future experimental measurements, we } decompose the $F^{\mu\nu}$ in Eqs. (\ref{eq:j_2_mu_v}, \ref{eq:j_2_mu_5})
by using the Eq. (\ref{eq:decompose_F}) and observing them in fluid rest frame
$u^\mu=(1,0,0,0)$.  After such decomposition, the vector and axial currents in fluid rest frame are given by
\begin{eqnarray}
n^{(2)} & = & \frac{\mu}{2\pi^{2}}\boldsymbol{\omega}^{2}+\left(\frac{1}{6}\kappa^{\epsilon}+\frac{5}{12\pi^{2}}\right)\boldsymbol{\varepsilon}\cdot\mathbf{E}-\left(\frac{1}{3}\kappa^{\epsilon}+\frac{1}{4\pi^{2}}\right)\boldsymbol{\omega}\cdot\mathbf{B}\nonumber \\
 &  & +\frac{C}{12\pi^{2}}(\mathbf{E}^{2}+\mathbf{B}^{2})+\left(\frac{1}{6}\kappa^{\epsilon}+\frac{1}{4\pi^{2}}\right)\boldsymbol{\nabla}\cdot\mathbf{E},\label{eq:vector_jq}\\
\mathbf{j}^{(2)} & = & \frac{\mu}{6\pi^{2}}\boldsymbol{\varepsilon}\times\boldsymbol{\omega}-\left(\frac{1}{6}\kappa^{\epsilon}+\frac{1}{12\pi^{2}}\right)\mathbf{E}\times\boldsymbol{\omega}-\frac{1}{6}\kappa^{\epsilon}\boldsymbol{\varepsilon}\times\mathbf{B}+\frac{C}{6\pi^{2}}\mathbf{E}\times\mathbf{B}\nonumber \\
 &  & -\left(\frac{1}{6}\kappa^{\epsilon}+\frac{1}{6\pi^{2}}\right)\partial_{t}\mathbf{E}+\left(\frac{1}{6}\kappa^{\epsilon}+\frac{1}{6\pi^{2}}\right)\boldsymbol{\nabla}\times\mathbf{B},\label{eq:vector_j-1}\\
n_{5}^{(2)} & = & \frac{\mu_{5}}{2\pi^{2}}\boldsymbol{\omega}^{2}+\frac{1}{6}\kappa_{5}\boldsymbol{\varepsilon}\cdot\mathbf{E}-\frac{1}{3}\kappa_{5}\boldsymbol{\omega}\cdot\mathbf{B}+\frac{C_{5}}{12\pi^{2}}(\mathbf{E}^{2}+\mathbf{B}^{2})+\frac{1}{6}\kappa_{5}\boldsymbol{\nabla}\cdot\mathbf{E}\label{eq:axial_jq}\\
\mathbf{j}_{5}^{(2)} & = & \frac{\mu_{5}}{6\pi^{2}}\boldsymbol{\varepsilon}\times\boldsymbol{\omega}-\frac{1}{6}\kappa_{5}\mathbf{E}\times\boldsymbol{\omega}-\frac{1}{6}\kappa_{5}\boldsymbol{\varepsilon}\times\mathbf{B}-\frac{C_{5}}{6\pi^{2}}\mathbf{E}\times\mathbf{B}\nonumber \\
 &  & -\frac{1}{6}\kappa_{5}\partial_{t}\mathbf{E}+\frac{1}{6}\kappa_{5}\boldsymbol{\nabla}\times\mathbf{B},\label{eq:axial_j-1}
\end{eqnarray}
where $n^{(2)},n_{5}^{(2)}$ are time components of vector and
axial currents and $\mathbf{j}^{(2)},\mathbf{j}_{5}^{(2)}$ are space
components of them. We find that the time components of vector and
axial currents are modified by divergence of electric field $\boldsymbol{\nabla}\cdot\mathbf{E}$
and the space components of vector currents and axial currents are
modified by the time derivative of electric field $\partial_{t}\mathbf{E}$
and curl of magnetic field $\boldsymbol{\nabla}\times\mathbf{B}$.

\section{energy momentum tensor\label{sec:energy-momentum-tensor}}

The canonical energy-momentum tensor can be derived from the vector Wigner function
by
\begin{eqnarray}
T^{\mu\nu} & = & \int d^{4}p\mathcal{V}^{\mu}p^{\nu}=\int d^{4}p(\mathcal{J}_{+1}^{\mu}+\mathcal{J}_{-1}^{\mu})p^{\nu}.
\end{eqnarray}
Following the same procedure as in Ref.~\citep{Yang:2020mtz}, we obtain the
second-order energy-momentum tensor
\begin{eqnarray}
T^{(2)\mu\nu} & = & T_{\textrm{vv}}^{(2)\mu\nu}+T_{\textrm{ve}}^{(2)\mu\nu}+T_{\textrm{ee}}^{(2)\mu\nu}+\Delta^{\textrm{\tiny{I}}}T^{(2)\mu\nu}+\Delta^{\textrm{\tiny{II}}}T^{(2)\mu\nu},\label{eq:T2_c}
\end{eqnarray}
where $T_{\textrm{vv}}^{(2)\mu\nu},T_{\textrm{ve}}^{(2)\mu\nu},T_{\textrm{ee}}^{(2)\mu\nu}$
are coupling terms of vorticity-vorticity, vorticity-EM-field, EM-field-EM-field,
and have been already derived in Refs.~\citep{Yang:2020mtz,Hidaka:2022dmn},
\begin{eqnarray}
T_{\textrm{vv}}^{(2)\mu\nu} & = & -\frac{1}{12\pi^{2}}[\pi^{2}T^{2}+3(\mu^{2}+\mu_{5}^{2})][3u^{\mu}u^{\nu}(\omega^{2}+\varepsilon^{2})-\Delta^{\mu\nu}(\omega^{2}+\varepsilon^{2})\nonumber \\
 &  & -2(u^{\mu}\epsilon^{\nu\alpha\beta\gamma}+u^{\nu}\epsilon^{\mu\alpha\beta\gamma})u_{\alpha}\varepsilon_{\beta}\omega_{\gamma}-2(u^{\mu}\epsilon^{\nu\alpha\beta\gamma}-u^{\nu}\epsilon^{\mu\alpha\beta\gamma})u_{\alpha}\varepsilon_{\beta}\omega_{\gamma}],\\
T_{\textrm{ve}}^{(2)\mu\nu} & = & -\frac{\mu}{8\pi^{2}}[2u^{\mu}u^{\nu}(\omega\cdot B+\varepsilon\cdot E)-(\omega^{\mu}B^{\nu}+E^{\mu}\varepsilon^{\nu})-(\omega^{\nu}B^{\mu}+E^{\nu}\varepsilon^{\mu})\nonumber \\
 &  & -(u^{\mu}\epsilon^{\nu\alpha\beta\gamma}+u^{\nu}\epsilon^{\mu\alpha\beta\gamma})u_{\alpha}(E_{\beta}\omega_{\gamma}+\varepsilon_{\beta}B_{\gamma})-2(u^{\mu}\epsilon^{\nu\alpha\beta\gamma}-u^{\nu}\epsilon^{\mu\alpha\beta\gamma})u_{\alpha}(E_{\beta}\omega_{\gamma}+\varepsilon_{\beta}B_{\gamma})],\\
T_{\textrm{ee}}^{(2)\mu\nu} & = & -\frac{1}{6}\kappa^{\epsilon}\left(\frac{1}{4}g^{\mu\nu}F_{\gamma\beta}F^{\gamma\beta}-F^{\gamma\mu}F_{\gamma}^{\ \nu}\right)+\frac{1}{24\pi^{2}}[u^{\mu}u^{\nu}E^{2}-\Delta^{\mu\nu}(E^{2}+2B^{2})\nonumber \\
 &  & +4(E^{\mu}E^{\nu}+B^{\mu}B^{\nu})+3(u^{\mu}\epsilon^{\nu\alpha\beta\gamma}+u^{\nu}\epsilon^{\mu\alpha\beta\gamma})u_{\alpha}E_{\beta}B_{\gamma}\nonumber \\
 &  & +3(u^{\mu}\epsilon^{\nu\alpha\beta\gamma}-u^{\nu}\epsilon^{\mu\alpha\beta\gamma})u_{\alpha}E_{\beta}B_{\gamma}].\label{eq:T2_ee}
\end{eqnarray}
The additional terms $\Delta^{\textrm{\tiny{I}}}T^{(2)\mu\nu}$ and
$\Delta^{\textrm{\tiny{II}}}T^{(2)\mu\nu}$ are new. The term $\Delta^{\textrm{\tiny{I}}}T^{(2)\mu\nu}$
comes from $\Delta f^{(2)}$ in Eq.(\ref{eq:J_2_result-tf0}) and
reads
\begin{eqnarray}
\Delta^{\textrm{\tiny{I}}}T^{(2)\mu\nu} & = & \frac{1}{9\pi^{2}}[\pi^{2}T^{2}+3(\mu^{2}+\mu_{5}^{2})]\left[\left(3u^{\mu}u^{\nu}-\Delta^{\mu\nu}\right)\varepsilon^{2}+\left(u^{\mu}\epsilon^{\nu\lambda\alpha\beta}+u^{\nu}\epsilon^{\mu\lambda\alpha\beta}\right)\varepsilon_{\lambda}u_{\alpha}\omega_{\beta}\right]\nonumber \\
 &  & +\frac{\mu}{12\pi^{2}}(3u^{\mu}u^{\nu}-\Delta^{\mu\nu})\varepsilon\cdot E.
\end{eqnarray}
The term $\Delta^{\textrm{\tiny{II}}}T^{(2)\mu\nu}$ is from the contribution
$\Delta{\mathcal{J}}_{\mu}^{(2)}$ with only derivative terms on the
EM field involved
\begin{eqnarray}
    \Delta^{\textrm{\tiny{II}}}T^{(2)\mu\nu} & = & \frac{\mu}{24\pi^{2}}\left(u^{\mu}\partial_{\lambda}F^{\lambda\nu}+u^{\nu}\partial_{\lambda}F^{\lambda\mu}\right)+\frac{\mu}{12\pi^{2}}\left(u^{\nu}u^{\mu}-\Delta^{\mu\nu}\right)u^{\rho}\partial^{\lambda}F_{\lambda\rho} +\frac{\mu}{24\pi^{2}}u_{\lambda}\left(\partial^{\mu}F^{\nu\lambda}+\partial^{\nu}F^{\mu\lambda}\right) \nonumber \\
     &  & +\frac{\mu}{12\pi^{2}}u^{\lambda}u_{\rho}\left(u^{\mu}\partial_{\lambda}F^{\nu\rho}+u^{\nu}\partial_{\lambda}F^{\mu\rho}\right)
      -\frac{\mu}{8\pi^{2}}\left(u^{\mu}\partial_{\lambda}F^{\lambda\nu}-u^{\nu}\partial_{\lambda}F^{\lambda\mu}\right).\label{eq:T_2}
    \end{eqnarray}
Similar to the results of vector and axial currents, the second-order
energy-momentum tensor also receives extra contributions from the
varying EM field in space or time.
{The terms in the first line of right hand side in Eq.~(\ref{eq:T_2}) are consistent with the results obtained in Ref.~\citep{Mameda:2023ueq}. While, the terms in the second line of Eq.~(\ref{eq:T_2}) are new.}
We note that the derivative terms on the EM field
in energy-momentum are all finite and receive no divergence contribution.

Moreover, we have also verified that our results of currents and energy
momentum tensor satisfy the following currents and energy momentum
conservation law up to second order
\begin{eqnarray}
\partial_{\mu}j^{\mu} & = & 0,\ \partial_{\mu}j_{5}^{\mu}=-\frac{1}{2\pi^{2}}E\cdot B,\ \partial_{\mu}T^{\mu\nu}=F^{\nu\rho}j_{\rho}.
\end{eqnarray}


\section{Corrections to the spin polarization }\label{sec:polarization}

In this section, we implement the currents in the phase space given
in Eq. (\ref{eq:J_2_result-tf0}) to compute the corrections from
the space-time dependent EM fields to the spin polarization pseudo
vector for the $\Lambda$ and $\overline{\Lambda}$ hyperons. The
spin polarization pseudo vector $\mathcal{S}^{\mu}$ can be derived
by the modified Cooper-Frye formula \citep{Becattini:2013fla,Fang:2016uds},
\begin{equation}
\mathcal{S}^{\mu}=\frac{\int d\Sigma\cdot p\mathcal{\widetilde{J}}_{5}^{\mu}}{2m\int d\Sigma\cdot\mathcal{\widetilde{N}}},\label{eq:CS}
\end{equation}
where $\mathcal{\widetilde{J}}_{5}^{\mu}$ and $\mathcal{\widetilde{N}}$
are the axial current and particle number density in phase space and $\Sigma^{\mu}$ is the (chemical) freeze out hypersurface.
Following Refs.~\citep{Yi:2021ryh,Yi:2021unq,Wu:2022mkr,Yi:2023tgg},
we assume that the Wigner function for the massless fermions can be
extended to the massive cases and write down the expression for $\mathcal{\widetilde{J}}_{5}^{\mu}$
and $\mathcal{\widetilde{N}}^\mu$ in terms of Wigner functions,
\begin{eqnarray}
\mathcal{\mathcal{\widetilde{J}}}_{5}^{\mu} & \equiv & p_u \int d p_u \, \theta(p_u)\left(\mathcal{J}_{+}^{\mu}-\mathcal{J}_{-}^{\mu}\right),\nonumber \\
\mathcal{\widetilde{N}}^{\mu} &  \equiv & p_u \int dp_u\, \theta(p_u)\left(\mathcal{J}_{+}^{\mu}+\mathcal{J}_{-}^{\mu}\right),
\end{eqnarray}
where $p_u=u\cdot p$ is free variable inside the integral and  $p_u=\sqrt{m^2 - \Delta^{\mu\nu}p_\mu p_\nu}$ has been on-shell when it is outside the integral.   We choose the fluid comoving frame in this section.
The $m$ in the denominator is the mass of hyperon. The $\theta(p_u)$ means that we focus on the
contribution from particles and neglect the anti-particles for simplicity.

In the following, it is convenient to introduce the vector and axial charge distribution at zeroth order
functions,
\begin{equation}
f_{V}^{(0)}\equiv\frac{1}{2}(f_{+}^{(0)}+f_{-}^{(0)}),\quad f_{5}\equiv\frac{1}{2}(f_{+}^{(0)}-f_{-}^{(0)}).
\end{equation}
and the particle number at the freeze
out hypersurface $N$,
\begin{eqnarray}
N &=& \int d\Sigma^\sigma \mathcal{\widetilde{N}}_\sigma
\equiv  N^{(0)} + N^{(1)} + N^{(2)}
\end{eqnarray}
with the superscripts $(n)$ corresponding to the expansions  for Wigner functions defined in previous sections,
\begin{eqnarray}
N^{(0)} =\int d\Sigma^{\sigma}p_{\sigma}f_{V}^{(0)},\ \ \ 
N^{(1)} = \int d\Sigma^{\sigma}\tilde{ \mathcal{N}}_\sigma^{(1)},\ \ \ 
N^{(2)} = \int d\Sigma^{\sigma} \tilde{ \mathcal{N}}_\sigma^{(2)}
.\label{eq:particle_number}
\end{eqnarray}

In Refs. \citep{Yi:2021ryh,Yi:2021unq,Wu:2022mkr,Yi:2023tgg}, the
authors have set $f_{+}^{(0)}=f_{-}^{(0)}$ for simplicity, which leads to $\mathcal{J}_{+,\sigma}^{(1)}+\mathcal{J}_{-,\sigma}^{(1)}=0$.
Interesting, we find that there is second order corrections to particle
number from EM fields. Here, we set $\tilde{f}^{(2)}=0$ for simplicity
and will discuss the influence of nonzero $\tilde{f}^{(2)}$ in Sec.
\ref{sec:general-solution}.

Recovering the $s$ dependence in Eqs. (\ref{eq:j2_5c}) and inserting
the expression for $\mathcal{J}_{\pm}^{\mu}$ into Eq. (\ref{eq:CS}),
we can decompose $\mathcal{S}^{\mu}$ into four parts
\begin{equation}
\mathcal{S}^{\mu}=\mathcal{S}^{(0)\mu}+\mathcal{S}^{(1)\mu}+\mathcal{S}^{(2)\mu}+\mathcal{S}_{\partial,\textrm{EM}}^{\mu}.
\end{equation}
The zeroth-order  $\mathcal{S}^{(0)\mu}$ reads,
\begin{equation}
\mathcal{S}^{(0)\mu}=\frac{1}{2mN^{(0)}}\int d\Sigma^{\sigma}p_{\sigma}p^{\mu}f_{5}^{(0)}. \label{eq:S(0)}
\end{equation}
Since $f_{5}^{(0)}\propto\mu_{5}$ when $\mu_{5}\rightarrow0$, the above
term is related to the polarization induced by the axial charge in
the final state \citep{Gao:2021rom}. 
The first-order $\mathcal{S}^{(1)\mu}$ stands for the ordinary terms up to
$\mathcal{O}(\hbar)$ for the spin polarization pseudo vector,
\begin{eqnarray}
 \mathcal{S}^{(1)\mu}& = & \mathcal{S}_{\textrm{thermal}}^{\mu}+\mathcal{S}_{\textrm{shear}}^{\mu}+\mathcal{S}_{\textrm{accT}}^{\mu}+\mathcal{S}_{\textrm{chemical}}^{\mu}+\mathcal{S}_{\textrm{EM}}^{\mu},
\end{eqnarray}
where $\mathcal{S}_{\textrm{thermal}}^{\mu},\mathcal{S}_{\textrm{shear}}^{\mu},\mathcal{S}_{\textrm{accT}}^{\mu},\mathcal{S}_{\textrm{chemical}}^{\mu}$
and $\mathcal{S}_{\textrm{EM}}^{\mu}$ denote  the polarizations
induced by the thermal vorticity, shear viscous tensor, fluid acceleration,
$\nabla(\mu/T)$ and EM fields in the $\mathcal{O}(\hbar)$, respectively.
The detailed expressions for these terms can be found  in Refs.
\citep{Yi:2021ryh,Yi:2021unq,Wu:2022mkr,Yi:2023tgg}. 
For second-order contribution  $\mathcal{S}^{(2)\mu}$, we will only focus on the  contribution directly from the numerator $\tilde{\mathcal{J}}_5^{(2)\mu}$  and neglect the trival contribution  from the denominator  $\tilde{\mathcal{N}}^{(1)} $ and $\tilde{\mathcal{N}}^{(2)} $. The polarization induced
by the second order effects with constant EM fields,
\begin{eqnarray}
\mathcal{S}^{(2)\mu} & = & \frac{p_{u}}{mN^{(0)}}\int d\Sigma\cdot p[\Omega_{\ \rho}^{\nu}\Omega^{\lambda\rho}X_{\mu\nu\lambda}^{\Omega\Omega}
+(F_{\ \rho}^{\nu}\Omega^{\lambda\rho}+F_{\ \rho}^{\lambda}\Omega^{\nu\rho})X_{\mu\nu\lambda}^{F\Omega}+F_{\ \rho}^{\nu}F^{\lambda\rho}X_{\mu\nu\lambda}^{FF}], \label{eq:S(2)}
\end{eqnarray}
where the coefficients $X_{\mu\nu\lambda}^{\Omega\Omega}$, $X_{\mu\nu\lambda}^{F\Omega}$ and $ X_{\mu\nu\lambda}^{FF}$ are given by
\begin{eqnarray}
    X_{\mu\nu\lambda}^{\Omega\Omega}	&=&	-\frac{1}{16p_{u}^{2}}[p_{u}(g_{\mu\nu}p_{\lambda}+g_{\mu\lambda}p_{\nu})-u_{\mu}p_{\nu}p_{\lambda}-p_{\mu}(u_{\nu}p_{\lambda}+u_{\lambda}p_{\nu})]f_5^{(0)\prime\prime}\nonumber\\
& &    -\frac{1}{16p_{u}^{3}}p_{\mu}p_{\nu}p_{\lambda}(f^{(0)\prime\prime}-\beta p_{u}f_5^{(0)\prime\prime\prime}),\\
    X_{\mu\nu\lambda}^{F\Omega}	&=&	\frac{1}{16p_{u}^{3}}(g_{\mu\nu}p_{\lambda}+g_{\mu\lambda}p_{\nu})(f_5^{(0)\prime}-2\beta p_{u}f_5^{(0)\prime\prime})-\frac{1}{32p_{u}^{4}}u_{\mu}p_{\nu}p_{\lambda}(3f_5^{(0)\prime}-2\beta p_{u}f_5^{(0)\prime\prime})\nonumber\\
& &            -\frac{1}{16p_{u}^{4}}p_{u}(u_{\nu}p_{\lambda}+u_{\lambda}p_{\nu})(3f_5^{(0)\prime}-\beta p_{u}f_5^{(0)\prime\prime})
-\frac{1}{32p_{u}^{5}}p_{\mu}p_{\nu}p_{\lambda}(3f_5^{(0)\prime}-3\beta p_{u}f_5^{(0)\prime\prime}+\beta^{2}p_{u}^{2}f_5^{(0)\prime\prime\prime})\nonumber\\
& &            +\frac{1}{16p_{u}^{3}}[u_{\mu}(u_{\nu}p_{\lambda}+u_{\lambda}p_{\nu})+p_{\mu}u_{\nu}u_{\lambda}-p_{u}(g_{\mu\nu}u_{\lambda}+g_{\mu\lambda}u_{\nu})]
f_5^{(0)\prime},\\
    X_{\mu\nu\lambda}^{FF}	&=&	-\frac{1}{16p_{u}^{5}}(g_{\mu\nu}p_{\lambda}+g_{\mu\lambda}p_{\nu})
    (3f_5^{(0)}-3\beta p_{u}f_5^{(0)\prime}+\beta^{2}p_{u}^{2}f_5^{(0)\prime\prime})+\frac{1}{8p_{u}^{4}}(g_{\mu\nu}u_{\lambda}+g_{\mu\lambda}u_{\nu})
    (3f_5^{(0)}-\beta p_{u}f_5^{(0)\prime})\nonumber\\
& &   +\frac{1}{48p_{u}^{6}}[p_{\mu}(u_{\nu}p_{\lambda}+u_{\lambda}p_{\nu})+u_{\mu}p_{\nu}p_{\lambda}]
(15f_5^{(0)}-12\beta p_{u}f_5^{(0)\prime}+3\beta^{2}p_{u}^{2}f_5^{(0)\prime\prime})\nonumber\\
& & -\frac{1}{8p_{u}^{3}}[u_{\mu}(u_{\nu}p_{\lambda}+u_{\lambda}p_{\nu})+p_{\mu}u_{\nu}u_{\lambda}]
(2f_5^{(0)}-\beta p_{u}f_5^{(0)\prime})+\frac{1}{8p_{u}^{4}}u_{\mu}u_{\nu}u_{\lambda}f_5^{(0)}\nonumber\\
& &            -\frac{1}{48p_{u}^{7}}p_{\mu}p_{\nu}p_{\lambda}
(15f_5^{(0)}-15\beta p_{u}f_5^{(0)\prime}+6\beta^{2}p_{u}^{2}f_5^{(0)\prime\prime}-\beta^{3}p_{u}^{3}f_5^{(0)\prime\prime\prime})
\end{eqnarray}
The coefficient $ X_{\mu\nu\lambda}^{F\Omega}$ represents  the coupling between the EM fields and thermal vortical
tensor.

The last term $\mathcal{S}_{\partial,\textrm{EM}}^{\mu}$ is proportional
to the derivative of EM fields,
\begin{equation}
\mathcal{S}_{\partial,\textrm{EM}}^{\mu}=\frac{p_{u}}{mN^{(0)}}\int d\Sigma\cdot p[(\partial_{\lambda}F^{\lambda\nu})X_{\mu\nu}^{\partial F}+\beta_{\lambda}(\partial^{\lambda}F^{\rho\nu})X_{\mu\rho\nu}^{\partial F}
+(\partial^{\rho}F^{\nu\lambda}+\partial^{\nu}F^{\rho\lambda})X_{\mu\rho\nu\lambda}^{\partial F}], \label{eq:SEM2nd}
\end{equation}
where the coefficients $ X_{\mu\nu}^{\partial F}	$, $X_{\mu\rho\nu}^{\partial F}$ and $X_{\mu\rho\nu\lambda}^{\partial F} $ are given by
\begin{eqnarray}
    X_{\mu\nu}^{\partial F}	&=&	\frac{1}{6p_{u}^{3}}g_{\mu\nu}(f_5^{(0)}-\beta p_{u}f_5^{(0)\prime})
    +\frac{1}{24p_{u}^{5}}p_{\mu}p_{\nu}(3f_5^{(0)}-3\beta p_{u}f_5^{(0)\prime}+\beta^{2}p_{u}^{2}f_5^{(0)\prime\prime})\nonumber\\
& &    -\frac{1}{24p_{u}^{4}}(u_{\mu}p_{\nu}+u_{\nu}p_{\mu})(3f_5^{(0)}-2\beta p_{u}f_5^{(0)\prime})
    +\frac{1}{48p_{u}^{3}}u_{\mu}u_{\nu}(4f_5^{(0)}+\beta p_{u}f_5^{(0)\prime})\nonumber\\
& &    -\frac{\beta}{48p_{u}^{3}}u_{\nu}p_{\mu}(f_5^{(0)\prime}-\beta p_{u}f_5^{(0)\prime\prime}),\\
X_{\mu\rho\nu}^{\partial F}	&=&	-\frac{1}{32p_{u}^{3}}(g_{\mu\rho}p_{\nu}-g_{\mu\nu}p_{\rho})
(f_5^{(0)\prime}-\beta p_{u}f_5^{(0)\prime\prime})
+\frac{1}{96p_{u}^{2}}(g_{\mu\rho}u_{\nu}-g_{\mu\nu}u_{\rho})(3f_5^{(0)\prime}-2\beta p_{u}f_5^{(0)\prime\prime})\nonumber\\
& &    +\frac{1}{48p_{u}^{3}}p_{\mu}(\beta_{\rho}p_{\nu}-\beta_{\nu}p_{\rho})(f_5^{(0)\prime\prime}-\beta p_{u}f_5^{(0)\prime\prime\prime})
-\frac{1}{48p_{u}^{2}}u_{\mu}(\beta_{\rho}p_{\nu}-\beta_{\nu}p_{\rho})f_5^{(0)\prime\prime},\\
X_{\mu\rho\nu\lambda}^{\partial F}	&=&	\frac{1}{48p_{u}^{5}}g_{\mu\lambda}p_{\rho}p_{\nu}
(3f_5^{(0)}-3\beta p_{u}f_5^{(0)\prime}+\beta^{2}p_{u}^{2}f_5^{(0)\prime\prime})
-\frac{1}{16p_{u}^{4}}g_{\mu\lambda}(u_{\rho}p_{\nu}+u_{\nu}p_{\rho})f_5^{(0)}\nonumber\\
& &+\frac{1}{24p_{u}^{3}}g_{\mu\lambda}u_{\rho}u_{\nu}f_5^{(0)}
    -\frac{\beta}{96p_{u}^{3}}u_{\lambda}(g_{\mu\rho}p_{\nu}+g_{\mu\nu}p_{\rho})
    (f_5^{(0)\prime}-\beta p_{u}f_5^{(0)\prime\prime})
    +\frac{\beta}{96p_{u}^{2}}u_{\lambda}(g_{\mu\rho}u_{\nu}+g_{\mu\nu}u_{\rho})f_5^{(0)\prime}\nonumber\\
& &    +\frac{\beta}{24p_{u}^{3}}g_{\mu\lambda}(u_{\rho}p_{\nu}+u_{\nu}p_{\rho})f_5^{(0)\prime}
    -\frac{\beta}{96p_{u}^{5}}u_{\lambda}p_{\mu}p_{\nu}p_{\rho}(3f_5^{(0)\prime}-3\beta p_{u}f_5^{(0)\prime\prime}+\beta^{2}p_{u}^{2}f_5^{(0)\prime\prime\prime})\nonumber\\
& &    +\frac{\beta}{96p_{u}^{4}}u_{\mu}u_{\lambda}p_{\nu}p_{\rho}(3f_5^{(0)\prime}-2\beta p_{u}f_5^{(0)\prime\prime})
    -\frac{\beta}{48p_{u}^{3}}u_{\mu}u_{\lambda}(u_{\nu}p_{\rho}+u_{\rho}p_{\nu})f_5^{(0)\prime}\nonumber\\
& &    +\frac{\beta}{96p_{u}^{4}}u_{\lambda}p_{\mu}(u_{\nu}p_{\rho}+u_{\rho}p_{\nu})(3f_5^{(0)\prime}-2\beta p_{u}f_5^{(0)\prime\prime})
\end{eqnarray}

We find that both $\mathcal{S}^{(2)\mu}$ and $\mathcal{S}_{\partial,\textrm{EM}}^{\mu}$
are proportional to $\mu_{5}$ in the small $\mu_{5}$ limit. For
massive cases, the $\mu_{5}$ may be simply replaced by the spin chemical
potential. The spin hydrodynamics contain the non-hydrodynamic modes
\citep{Hattori:2019lfp,Fukushima:2020ucl,Hongo:2021ona,Daher:2022wzf,Xie:2023gbo,Ren:2024pur}.
Thus, the spin chemical potential is considered to decay much more
rapidly than other ordinary hydrodynamic quantities, e.g. in Refs.
\citep{Wang:2021ngp,Wang:2021wqq} the authors found that both spin
chemical potential and spin density decay exponentially. However,
a very recent study has observed attractors in spin hydrodynamics
at late times \citep{Wang:2024afv}. This result indicates that the
spin chemical potential or spin density can be treated as ordinary
hydrodynamic quantities. Therefore, the corrections $\mathcal{S}^{(2)\mu}$
and $\mathcal{S}_{\partial,\textrm{EM}}^{\mu}$ may also need to be
considered in the future numerical simulations.


\section{general solution for $f^{(1)}$ and $\tilde{f}^{(2)}$}
\label{sec:general-solution} In Sec.~\ref{sec:Wigner-function-at},
we have taken the solution that $f^{(1)}=\tilde{f}^{(2)}=0$ which
satisfies the Eqs. (\ref{eq:evolution_1_part_1}) and (\ref{eq:evolution_2_part_2}).
To obtain the general solution, we construct all possible terms which
satisfy Lorentz invariance as well as charge and parity invariance
for $f^{(1)}$ and $\tilde{f}^{(2)}$. Taking into account the independence
of various terms by the Eq. (\ref{eq:partial_t_E}), we can write
the general expressions as
\begin{eqnarray}
f^{(1)} & = & (\omega\cdot p)\beta^{2}\mathcal{X}_{\omega}^{(1)}+(B\cdot p)\beta^{3}\mathcal{X}_{B}^{(1)},\label{f1-g}\\
\tilde{f}^{(2)} & = & \mathcal{X}_{\Omega\Omega}+\mathcal{X}_{\Omega F}+\mathcal{X}_{FF}+\mathcal{X}_{\partial F},\label{f2-g}
\end{eqnarray}
where $\tilde{f}^{(2)}$ have been categorized into four parts
\begin{eqnarray}
\mathcal{X}_{\Omega\Omega} & = & \omega^{2}\beta^{2}\mathcal{X}_{\omega\omega1}^{(2)}+\varepsilon^{2}\beta^{2}\mathcal{X}_{\varepsilon\varepsilon1}^{(2)}+(\omega\cdot p)^{2}\beta^{4}\mathcal{X}_{\omega\omega2}^{(2)}\nonumber \\
 &  & +(\varepsilon\cdot p)^{2}\beta^{4}\mathcal{X}_{\varepsilon\varepsilon2}^{(2)}+\epsilon^{\nu\lambda\rho\sigma}u_{\nu}p_{\lambda}\omega_{\rho}\varepsilon_{\sigma}\beta^{3}\mathcal{X}_{\omega\varepsilon}^{(2)},\\
\mathcal{X}_{\Omega F} & = & (\omega\cdot B)\beta^{3}\mathcal{X}_{\omega B1}^{(2)}+(\varepsilon\cdot E)\beta^{3}\mathcal{X}_{\varepsilon E1}^{(2)}+(\omega\cdot p)(B\cdot p)\beta^{5}\mathcal{X}_{\omega B2}^{(2)}\nonumber \\
 &  & +(\varepsilon\cdot p)(E\cdot p)\beta^{5}\mathcal{X}_{\varepsilon E2}^{(2)}+\epsilon^{\nu\lambda\rho\sigma}u_{\nu}p_{\lambda}\omega_{\rho}E_{\sigma}\beta^{4}\mathcal{X}_{\omega E}^{(2)}\nonumber \\
 &  & +\epsilon^{\nu\lambda\rho\sigma}u_{\nu}p_{\lambda}B_{\rho}\varepsilon_{\sigma}\beta^{4}\mathcal{X}_{\varepsilon B}^{(2)},\\
\mathcal{X}_{FF} & = & B^{2}\beta^{4}\mathcal{X}_{BB1}^{(2)}+E^{2}\beta^{4}\mathcal{X}_{EE1}^{(2)}+(B\cdot p)^{2}\beta^{6}\mathcal{X}_{BB2}^{(2)}\nonumber \\
 &  & +(E\cdot p)^{2}\beta^{6}\mathcal{X}_{EE2}^{(2)}+\epsilon^{\nu\lambda\rho\sigma}u_{\nu}p_{\lambda}B_{\rho}E_{\sigma}\beta^{5}\mathcal{X}_{BE}^{(2)},\\
\mathcal{X}_{\partial F} & = & (\partial\cdot E)\beta^{3}\mathcal{X}_{\partial E1}^{(2)}+[(p\cdot\partial)(E\cdot p)]\beta^{5}\mathcal{X}_{\partial E2}^{(2)}\nonumber \\
 &  & +\epsilon^{\nu\lambda\rho\sigma}u_{\nu}p_{\lambda}(\partial_{\rho}B_{\sigma})\beta^{4}\mathcal{X}_{\partial B}^{(2)}.
\end{eqnarray}
Without loss of generality, we can assume all $\mathcal{X}$ are functions
of
\begin{eqnarray}
z & = & \beta\cdot p-\bar{\mu}_{s},\ \tilde{z}=\beta\cdot p+\bar{\mu}_{s}.
\end{eqnarray}
According to the Eq. (\ref{eq:partial_t_E}) terms $(u\cdot\partial)(E\cdot p),p_u(\varepsilon\cdot E)$
and $\epsilon^{\nu\lambda\rho\sigma}u_{\nu}p_{\lambda}\omega_{\rho}E_{\sigma}$
are linearly dependent, here we have kept $p_u(\varepsilon\cdot E)$
and $\epsilon^{\nu\lambda\rho\sigma}u_{\nu}p_{\lambda}\omega_{\rho}E_{\sigma}$,
but no $(u\cdot\partial)(E\cdot p)$.

For the first-order expression (\ref{f1-g}), we substitute it into
Eq. (\ref{eq:evolution_1_part_1}) and obtain
\begin{eqnarray}
0 & = & p^{\mu}\delta(p^{2})\nabla_{\mu}f^{(1)}\nonumber \\
 & = & -2(\omega\cdot p)(E\cdot p)\beta^{3}\frac{\partial\mathcal{X}_{\omega}^{(1)}}{\partial\tilde{z}}\delta(p^{2})-2(B\cdot p)(E\cdot p)\beta^{4}\frac{\partial\mathcal{X}_{B}^{(1)}}{\partial\tilde{z}}\delta(p^{2})\nonumber \\
 &  & +p_{0}(\omega\cdot E)\beta^{2}\mathcal{X}_{\omega}^{(1)}\delta(p^{2})-\epsilon_{\mu\nu\alpha\beta}p^{\mu}\omega^{\nu}u^{\alpha}B^{\beta}\beta^{2}\mathcal{X}_{\omega}^{(1)}\delta(p^{2})\nonumber \\
 &  & +p^{\mu}p^{\nu}(\partial_{\mu}B_{\nu})\beta^{3}\mathcal{X}_{B}^{(1)}\delta(p^{2})+p_{0}E\cdot B\beta^{3}\mathcal{X}_{B}^{(1)}\delta(p^{2}).\label{eq:nabla_f1}
\end{eqnarray}
Every term in this equation is linearly independent and should vanish
simultaneously. The only possibility is
\begin{eqnarray}
\mathcal{X}_{\omega}^{(1)} & = & \mathcal{X}_{B}^{(1)}=0.
\end{eqnarray}
This indicates $f^{(1)}=0$ is the unique solution of Wigner equations
at first order at global equilibrium. It is interesting to note that,
if we turn off the EM field, the equation $p^{\mu}\delta(p^{2})\nabla_{\mu}f^{(1)}=0$
is satisfied automatically so that we cannot constrain the coefficient
$\mathcal{X}_{\omega}^{(1)}$. Hence we can constrain the distribution
function by kinetic equations more tightly by imposing the EM field.

As for second-order expression (\ref{f2-g}), substituting it into
Eq. (\ref{eq:evolution_2_part_2})
\begin{eqnarray}
0 & = & p^{\mu}\delta(p^{2})\nabla_{\mu}\tilde{f}^{(2)}=p^{\mu}\delta(p^{2})\nabla_{\mu}(\mathcal{X}_{\Omega\Omega}+\mathcal{X}_{\Omega F}+\mathcal{X}_{FF}+\mathcal{X}_{\partial F}).\label{eq:nabla_f2}
\end{eqnarray}
and collecting all the independent terms, we find Eq. (\ref{eq:nabla_f2})
holds only if
\begin{eqnarray}
\mathcal{X}_{\varepsilon\varepsilon2}^{(2)} & = & \mathcal{X}_{\omega\omega2}^{(2)}=\mathcal{X}_{\omega\varepsilon}^{(2)}=0,\\
\mathcal{X}_{\varepsilon E1}^{(2)} & = & \mathcal{X}_{\omega B1}^{(2)}=\mathcal{X}_{\varepsilon E2}^{(2)}=\mathcal{X}_{\omega B2}^{(2)}=\mathcal{X}_{\omega E}^{(2)}=\mathcal{X}_{\varepsilon B}^{(2)}=0,\\
\mathcal{X}_{EE1}^{(2)} & = & \mathcal{X}_{BB1}^{(2)}=\mathcal{X}_{EE2}^{(2)}=\mathcal{X}_{BB2}^{(2)}=\mathcal{X}_{BE}^{(2)}=0,\\
\mathcal{X}_{\partial E1}^{(2)} & = & \mathcal{X}_{\partial E2}^{(2)}=\mathcal{X}_{\partial B}^{(2)}=0,\\
\frac{\partial\mathcal{X}_{\varepsilon\varepsilon1}^{(2)}}{\partial\tilde{z}} & = & \frac{\partial\mathcal{X}_{\omega\omega1}^{(2)}}{\partial\tilde{z}}=0,\label{eq:partial_X}
\end{eqnarray}
together with
\begin{eqnarray}
\mathcal{X}_{\omega\omega1}^{(2)}+\mathcal{X}_{\varepsilon\varepsilon1}^{(2)} & = & 0.\label{eq:X_ww1}
\end{eqnarray}
Combining the Eqs. (\ref{eq:partial_X}) and (\ref{eq:X_ww1}), we
can obtain the general expression
\begin{eqnarray}
\mathcal{X}_{\varepsilon\varepsilon1}^{(2)} & = & -\mathcal{X}_{\omega\omega1}^{(2)}=a(z),
\end{eqnarray}
where $a(z)$ are unknown functions which only depend on $z$. The
second-order distribution function is given by
\begin{eqnarray}
\tilde{f}^{(2)} & = & (\varepsilon^{2}-\omega^{2})a(z). \label{eq:f_2}
\end{eqnarray}

Now, let us comment on $\tilde{f}^{(2)}$ in Eq.~(\ref{eq:f_2}).
{In Ref.~\citep{Yang:2020mtz}, the general expression for the second-order distribution function was analyzed under a constant EM field. The analysis revealed that the general expression  dependents on two arbitrary functions of $z$. In contrast, when extending the scenario to include varying EM fields in space and time, we find that the number of arbitrary functions in the general solution reduces from two to one.}

Integrating over four-momentum, we can obtain the extra contributions to current
and energy-momentum tensor with chirality $s$
\begin{eqnarray}
\tilde{\Delta}j_{s}^{\mu} & = & (\varepsilon^{2}-\omega^{2})u^{\mu}A_{j},\\
\tilde{\Delta}T_{s}^{\mu\nu} & = & (\varepsilon^{2}-\omega^{2})\left(u^{\mu}u^{\nu}-\frac{1}{3}\Delta^{\mu\nu}\right)A_{T},
\end{eqnarray}
where
\begin{eqnarray}
A_{j} & = & 2\pi\beta^{2}\int dp_{0}p_{0}^{2}a(\beta p_{0}-\bar{\mu}_{s})[\theta(p_{0})-\theta(-p_{0})],\\
A_{T} & = & 2\pi\beta^{2}\int dp_{0}p_{0}^{3}a(\beta p_{0}-\bar{\mu}_{s})[\theta(p_{0})-\theta(-p_{0})].
\end{eqnarray}
It is easy to verify that $A_{j}$ and $A_{T}$ satisfy the following
relation
\begin{eqnarray}
A_{j} & = & \frac{1}{3\beta}\frac{\partial(\beta^{2}A_{T})}{\partial\bar{\mu}_{s}}.
\end{eqnarray}
When $A_{T}=-(\pi^{2}T^{2}+3\mu_{s}^{2})/(16\pi^{2})$,
this result is also consistent to the result in Ref. \citep{Buzzegoli:2017cqy,Buzzegoli:2018wpy}.
{Similarly, we can get the correction the spin polarization pseudo-vector $\mathcal{S}^\mu$ from $\tilde{f^{(2)}}$,
\begin{equation}
    \mathcal{S}_{f(2)}^{\mu}=\frac{1}{2mN}\int d\Sigma^{\sigma}p_{\sigma}p^{\mu}[a(z)-a(\bar{z})],
\end{equation}
which may be merged into $\mathcal{S}_{(0)}^{\mu}$ in Eq.~(\ref{eq:S(0)}).
}

\section{Summary}
\label{sec:summary} Within the Wigner function formalism and semiclassical
expansion scheme, we solve the Wigner equation and obtain the Wigner
functions at global equilibrium with constant vorticity but space-time
dependent EM field up to second order. {The corrections from space-time dependent EM fields to Wigner functions are shown in Eq.~(\ref{eq:J_2_result-tf0}).
Different with Ref.~\citep{Mameda:2023ueq},} without introducing additional
constraint conditions, we must introduce an additional contribution
$\Delta f^{(2)}$ to the second-order distribution function $f^{(2)}$
in order to satisfy the Wigner equation. Such method of finding the
solution at global equilibrium can be generalized to any higher order.
Hence this work provide a systematic method to find the solution of
Wigner equation under space-time dependent EM field in global equilibrium.

The vector and axial currents and energy-momentum tensor can be obtained
from the appropriate moments of the Wigner function. We find that
the second-order currents and energy-momentum tensor receive additional
contributions from the varying EM field in space or time, {see Eqs.~(\ref{eq:j_2_mu_v}, \ref{eq:j_2_mu_5}, \ref{eq:T_2}) for those corrections.}  The terms
associated with $\partial_{\lambda}F^{\mu\lambda}$ can be related
to the background electric current by the Maxwell's equation, which
implies that background electric currents can directly induce the
vector and axial currents as well as energy-momentum tensor. In particular,
we have obtained the new terms associated with $u^{\lambda}\partial_{\lambda}F^{\mu\nu}$
which is absent in earlier works by imposing the additional constraint
conditions. We also find that the derivative terms have brought divergence
coefficients $\kappa^{\epsilon}$ in vector current while there is
no divergence in axial current or energy-momentum tensor. We have
checked that second order currents and energy momentum tensor satisfy
the currents conservation and energy momentum conservation.

{
We have also computed the corrections from EM fields to spin polarization pesudo-vector. The new corrections are $\mathcal{S}^\mu_{(0)}$ and $\mathcal{S}^\mu_{\partial,\textrm{EM}}$ shown in Eqs.~(\ref{eq:S(2)}) and (\ref{eq:SEM2nd}), respectively. Those new corrections need to be studied numerically in the future. }

We also have analyzed the general expressions for the distribution
function $f^{(1)}$ and $\tilde{f}^{(2)}$, which can not be determined
by the iterative relations of the Wigner equation. We find that the
space-time dependent electromagnetic field provides a tighter constraint
on the solutions of Wigner functions in global equilibrium compared
with constant electromagnetic field.

\acknowledgments{This work is supported in part by the National Key Research and Development Program of China under Contract No. 2022YFA1605500, by the Chinese Academy of Sciences (CAS) under Grants No. YSBR-088 and by National Natural Science Foundation of China (NSFC) under Grants No. 12075235, No. 12135011, No. 12175123,  and No. 12321005.}

\appendix

\section{Integrals associated with $\Delta f^{(2)}$ and $\Delta{\mathcal{J}}_{\mu}^{(2)}$}
 \label{sec:integral-for-currents}

In this appendix, we will present some details on the calculation of the integrals which comes from $\Delta f^{(2)}$ and $\Delta{\mathcal{J}}_{\mu}^{(2)}$. These terms include $\Delta^{\textrm{\tiny{I}}} j_{s}^{(2)\mu}$, $\Delta^{\textrm{\tiny{II}}} j_{s}^{(2)\mu}$ $\Delta^{\textrm{\tiny{I}}} T^{(2)\mu\nu}_s$, and $\Delta^{\textrm{\tiny{II}}} T^{(2)\mu\nu}_s$, which are absent in the previous work \citep{Yang:2020mtz}.
\begin{eqnarray}
\Delta^{\textrm{\tiny{I}}} j_{s}^{(2)\mu} & = & \int d^{4}pp^{\mu}\Delta f^{(2)}\delta(p^{2})\nonumber \\
 & = & -\frac{1}{24}u^{\mu}F_{\rho\lambda}\Omega_{\sigma}^{\ \lambda}\beta^{\rho}\beta^{\sigma}\int d^{4}pp_uf^{\prime\prime\prime}\delta(p^{2})-\frac{1}{12}u^{\mu}u^{\rho}\text{\ensuremath{\Omega}}_{\rho\lambda}\Omega_{\sigma}^{\ \lambda}\beta^{\sigma}\int d^{4}pp_u^{2}f^{\prime\prime\prime}\delta(p^{2})\nonumber \\
 &  & -\frac{1}{36}\Delta^{\mu\rho}\text{\ensuremath{\Omega}}_{\rho\lambda}\Omega_{\sigma}^{\ \lambda}\beta^{\sigma}\int d^{4}p\bar{p}^{2}f^{\prime\prime\prime}\delta(p^{2}).
\end{eqnarray}
\begin{eqnarray}
\Delta^{\textrm{\tiny{II}}}j_{s}^{(2)\mu} & = & \int d^{4}p \Delta{\mathcal{J}}^{(2)\mu} \nonumber \\
& = & -\frac{2}{3}(\partial^{\lambda}F_{\mu\lambda})\int d^{4-\epsilon}pf\delta^{\prime}(p^{2})
-\frac{1}{12}\beta_{\kappa}(\partial^{\kappa}F_{\mu\lambda})\beta^{\lambda}\int d^{4}pf^{\prime\prime}\delta(p^{2})\nonumber \\
 &  & -\frac{1}{12}(\partial^{\nu}F_{\nu\rho})\beta^{\rho}u_{\mu}\int d^{4}pp_uf^{\prime}\delta^{\prime}(p^{2})
 +\frac{1}{6} \beta^{\lambda}(\partial_{\lambda} F_{\mu\nu}) u^{\nu}\int d^{4}pp_uf^{\prime}\delta^{\prime}(p^{2})\nonumber \\
 &  & +\frac{1}{3}(\partial^{\lambda}F_{\lambda\nu}) u_{\mu}u^{\nu}\int d^{4-\epsilon}pp_u^{2}f\delta^{\prime\prime}(p^{2})
 +\frac{1}{3(3-\epsilon)}(\partial^{\lambda}F_{\lambda\nu})\Delta_{\mu}^{\ \nu}
 \int d^{4-\epsilon}p\bar{p}^{2}f\delta^{\prime\prime}(p^{2})\nonumber \\
 &  & -\frac{1}{3}(\partial^{\lambda}F_{\mu\nu}) u^{\nu}u_{\lambda}\int d^{4-\epsilon}pp_u^{2}f\delta^{\prime\prime}(p^{2})
 -\frac{1}{3(3-\epsilon)}(\partial^{\lambda}F_{\mu\nu})\Delta_{\;\lambda}^{\nu}
 \int d^{4-\epsilon}p\bar{p}^{2}f\delta^{\prime\prime}(p^{2})\nonumber \\
 &  & -\frac{1}{18}(\partial_{\lambda}F_{\nu\rho})\beta^{\rho}
 (u_{\mu}\Delta^{\lambda\nu}+u^{\lambda}\Delta_{\mu}^{\;\nu}+u^{\nu}\Delta_{\mu}^{\;\lambda})
 \int d^{4}pp_u\bar{p}^{2}f^{\prime}\delta^{\prime\prime}(p^{2})\nonumber\\
 &  & -\frac{1}{18}\beta^{\lambda}(\partial_{\lambda} F_{\nu\rho})\beta^{\rho}\Delta_{\mu}^{\;\nu}
 \int d^{4}p\bar{p}^{2}f^{\prime\prime}\delta^{\prime}(p^{2})
 \label{eq:j_2_integrate}
\end{eqnarray}
\begin{eqnarray}
 \Delta^{\textrm{\tiny{I}}}T_{s}^{(2)\mu\nu}  & = & \int d^{4}p p^{\mu}p^\nu\Delta f^{(2)}\delta(p^{2})\nonumber \\
 & = & -\frac{1}{24}u^{\mu}u^{\nu}F_{\rho\lambda}\Omega_{\sigma}^{\ \lambda}\beta^{\rho}\beta^{\sigma}\int d^{4}pp_u^{2}f^{\prime\prime\prime}\delta(p^{2})\nonumber \\
 &  & -\frac{1}{72}\Delta^{\mu\nu}F_{\rho\lambda}\Omega_{\sigma}^{\ \lambda}\beta^{\rho}\beta^{\sigma}\int d^{4}p\bar{p}^{2}f^{\prime\prime\prime}\delta(p^{2})\nonumber \\
 &  & -\frac{1}{12}u^{\mu}u^{\nu}u^{\rho}\text{\ensuremath{\Omega}}_{\rho\lambda}\Omega_{\sigma}^{\ \lambda}\beta^{\sigma}\int d^{4}pp_u^{3}f^{\prime\prime\prime}\delta(p^{2})\nonumber \\
 &  & -\frac{1}{36}(u^{\mu}\Delta^{\nu\rho}+u^{\nu}\Delta^{\mu\rho}+u^{\rho}\Delta^{\mu\nu})\text{\ensuremath{\Omega}}_{\rho\lambda}\Omega_{\sigma}^{\ \lambda}\beta^{\sigma}\int d^{4}pp_u\bar{p}^{2}f^{\prime\prime\prime}\delta(p^{2}).
\end{eqnarray}
\begin{eqnarray}
 \Delta^{\textrm{\tiny{II}}}T_{s}^{(2)\mu\nu} & = & \int d^{4}p \Delta\mathcal{J}_{s}^{\mu}p^{\nu}\nonumber \\
 & = &-\frac{2}{3}u^{\nu}(\partial_{\lambda}F^{\mu\lambda})\int d^{4}pp_uf\delta^{\prime}(p^{2})-\frac{1}{12}u^{\nu}\beta_{\kappa}(\partial^{\kappa}F^{\mu\lambda})\beta_{\lambda}\int d^{4}pp_uf^{\prime\prime}\delta(p^{2})\nonumber \\
 &  & -\frac{1}{12}u^{\mu}u^{\nu}(\partial^{\lambda}F_{\lambda\rho})\beta^{\rho}\int d^{4}pp_u^{2}f^{\prime}\delta^{\prime}(p^{2})-\frac{1}{36}\Delta^{\mu\nu}(\partial^{\lambda}F_{\lambda\rho})\beta^{\rho}\int d^{4}p\bar{p}^{2}f^{\prime}\delta^{\prime}(p^{2})\nonumber \\
 &  & +\frac{1}{6}\beta^{\lambda}(\partial_{\lambda}F^{\mu\rho})u^{\nu}u_{\rho}\int d^{4}pp_u^{2}f^{\prime}\delta^{\prime}(p^{2})+\frac{1}{18}\beta^{\lambda}(\partial_{\lambda}F^{\mu\rho})\Delta_{\rho}^{\nu}\int d^{4}p\bar{p}^{2}f^{\prime}\delta^{\prime}(p^{2})\nonumber \\
 &  & -\frac{1}{18}(\partial^{\mu}F_{\rho\lambda})\beta^{\lambda}\Delta^{\nu\rho}\int d^{4}p\bar{p}^{2}f^{\prime}\delta^{\prime}(p^{2})+\frac{1}{3}(\partial^{\lambda}F_{\lambda\rho})u^{\mu}u^{\nu}u^{\rho}\int d^{4}pp_u^{3}f\delta^{\prime\prime}(p^{2})\nonumber \\
 &  & +\frac{1}{9}(\partial^{\lambda}F_{\lambda\rho})(u^{\mu}\Delta^{\nu\rho}+u^{\nu}\Delta^{\mu\rho}+u^{\rho}\Delta^{\mu\nu})\int d^{4}pp_u\bar{p}^{2}f\delta^{\prime\prime}(p^{2})\nonumber \\
 &  & -\frac{1}{3}(\partial^{\lambda}F^{\mu\rho})u^{\nu}u_{\lambda}u_{\rho}\int d^{4}pp_u^{3}f\delta^{\prime\prime}(p^{2})\nonumber \\
 &  & -\frac{1}{9}(\partial^{\lambda}F^{\mu\rho})(u^{\nu}\Delta_{\lambda\rho}+u_{\lambda}\Delta_{\rho}^{\nu}+u_{\rho}\Delta_{\lambda}^{\nu})\int d^{4}pp_u\bar{p}^{2}f\delta^{\prime\prime}(p^{2})\nonumber \\
 &  & -\frac{1}{18}\beta^{\lambda}(\partial_{\lambda}F_{\sigma\rho})\beta^{\rho}(u^{\mu}\Delta^{\nu\sigma}+u^{\nu}\Delta^{\mu\sigma})\int d^{4}pp_u\bar{p}^{2}f^{\prime\prime}\delta^{\prime}(p^{2})\nonumber \\
 &  & -\frac{1}{18}(\partial_{\lambda}F_{\sigma\rho})\beta^{\rho}(u^{\mu}u^{\nu}\Delta^{\lambda\sigma}+u^{\mu}u^{\lambda}\Delta^{\nu\sigma}+u^{\nu}u^{\lambda}\Delta^{\mu\sigma})\int d^{4}pp_u^{2}\bar{p}^{2}f^{\prime}\delta^{\prime\prime}(p^{2})\nonumber \\
 &  & -\frac{1}{90}(\partial_{\lambda}F_{\sigma\rho})\beta^{\rho}(\Delta^{\mu\nu}\Delta^{\lambda\sigma}+\Delta^{\mu\lambda}\Delta^{\nu\sigma}+\Delta^{\mu\sigma}\Delta^{\nu\lambda})\int d^{4}p\bar{p}^{4}f^{\prime}\delta^{\prime\prime}(p^{2}),
\end{eqnarray}
Note that we have suppressed all the order indices  $(0)$ on the distribution function $f$ for brevity of symbol.
In deriving the above results, we  have used the following tensor decomposition
\begin{eqnarray}
\int d^{4}pp^{\lambda}Y & = & u^{\lambda}\int d^{4}pp_uY,\nonumber \\
\int d^{4}pp^{\mu}p^{\lambda}Y & = & u^{\mu}u^{\lambda}\int d^{4}pp_u^{2}Y+\frac{1}{3}\Delta^{\mu\nu}\int d^{4}p\bar{p}^{2}Y,\nonumber \\
\int d^{4}pp^{\mu}p^{\beta}p^{\lambda}Y & = & u^{\mu}u^{\beta}u^{\lambda}\int d^{4}pp_u^{3}Y+\frac{1}{3}(\Delta^{\mu\beta}u^{\lambda}+\Delta^{\mu\lambda}u^{\beta}+\Delta^{\lambda\beta}u^{\mu})\int d^{4}pp_u\bar{p}^{2}Y,\nonumber \\
\int d^{4}pp^{\mu}p^{\beta}p^{\lambda}p^{\sigma}Y & = & u^{\mu}u^{\beta}u^{\lambda}u^{\sigma}\int d^{4}pp_u^{4}Y+\frac{1}{15}(\Delta^{\mu\beta}\Delta^{\lambda\sigma}+\Delta^{\mu\lambda}\Delta^{\beta\sigma}+\Delta^{\mu\sigma}\Delta^{\beta\lambda})\int d^{4}p\bar{p}^{4}Y\nonumber \\
 &  & +\frac{1}{3}(\Delta^{\mu\beta}u^{\lambda}u^{\sigma}+\Delta^{\mu\lambda}u^{\beta}u^{\sigma}+\Delta^{\mu\sigma}u^{\beta}u^{\lambda}\nonumber \\
 &  & +\Delta^{\beta\lambda}u^{\mu}u^{\sigma}+\Delta^{\beta\sigma}u^{\mu}u^{\lambda}+\Delta^{\lambda\sigma}u^{\mu}u^{\beta})\int d^{4}pp_u^{2}\bar{p}^{2}Y,
\end{eqnarray}
for convergence coefficients, and
\begin{eqnarray}
\int d^{4-\epsilon}pp^{\mu}p^{\lambda}Y & = & u^{\mu}u^{\lambda}\int d^{4-\epsilon}pp_u^{2}Y+\frac{1}{3-\epsilon}\Delta^{\mu\nu}\int d^{4-\epsilon}p\bar{p}^{2}Y,
\end{eqnarray}
for divergence coefficients, where  $Y$ is function of $u\cdot p$ and $p^{2}$.

Here it would be sufficient to take  one convergence coefficient and one divergence
coefficient as an example.
\begin{eqnarray}
 \int d^{4}pp_uf^{\prime}\delta^{\prime}(p^{2}) & = & \int dp_{0}d|\vec{p}|p_{0}|\vec{p}|^{2}f^{\prime}\delta^{\prime}(p^{2})\nonumber\\
 & = & \pi\int dp_{0}d|\vec{p}||\vec{p}|f^{\prime}\left[\delta^{\prime}(p_{0}-|\vec{p}|)+\delta^{\prime}(p_{0}+|\vec{p}|)\right]\nonumber \\
 & = & \pi\int dp_{0}f^{\prime}\frac{\partial}{\partial p_{0}}\int d|\vec{p}||\vec{p}|\left[\delta(p_{0}-|\vec{p}|)+\delta(p_{0}+|\vec{p}|)\right]\nonumber \\
 & = & \pi\int dp_{0}f^{\prime}\frac{\partial}{\partial p_{0}}\{p_{0}[\theta(p_{0})-\theta(-p_{0})]\}\nonumber \\
 & = & \pi\int dp_{0}f^{\prime}[\theta(p_{0})-\theta(-p_{0})]
 \end{eqnarray}
 \begin{eqnarray}
 \int d^{4-\epsilon}pf\delta^{\prime}(p^{2}) & = & \frac{2\pi^{\frac{3-\epsilon}{2}}}{\Gamma\left(\frac{3-\epsilon}{2}\right)}\int dp_{0}d|\vec{p}||\vec{p}|^{2-\epsilon}f\delta^{\prime}(p^{2})\nonumber\\
 & = & \frac{\pi^{\frac{3-\epsilon}{2}}}{2\Gamma\left(\frac{3-\epsilon}{2}\right)}\int dp_{0}d|\vec{p}|p_{0}^{-1}|\vec{p}|^{1-\epsilon}f\left[\delta^{\prime}(p_{0}-|\vec{p}|)+\delta^{\prime}(p_{0}+|\vec{p}|)\right]\nonumber \\
 & = & \frac{\pi^{\frac{3-\epsilon}{2}}}{2\Gamma\left(\frac{3-\epsilon}{2}\right)}\int dp_{0}p_{0}^{-1}f\frac{\partial}{\partial p_{0}}\int d|\vec{p}||\vec{p}|^{1-\epsilon}\left[\delta(p_{0}-|\vec{p}|)+\delta(p_{0}+|\vec{p}|)\right]\nonumber \\
 & = & \frac{\pi^{\frac{3-\epsilon}{2}}}{2\Gamma\left(\frac{3-\epsilon}{2}\right)}\int dp_{0}p_{0}^{-1}f\frac{\partial}{\partial p_{0}}\{p_{0}^{1-\epsilon}\left[\theta(p_{0})-\theta(p_{0})\right]\}\nonumber \\
 & = & \frac{\pi^{\frac{3-\epsilon}{2}}}{2\Gamma\left(\frac{3-\epsilon}{2}\right)}\int dp_{0}(1-\epsilon)p_{0}^{-1-\epsilon}f\left[\theta(p_{0})-\theta(p_{0})\right].
\end{eqnarray}
where  we have calculate  the integrals in the fluid rest frame due to the Lorentz invariance.
Defining $y=p_{0}/T$, the integrals can written as
\begin{eqnarray}
 \int d^{4}pp_uf^{\prime}\delta^{\prime}(p^{2})
 & = & \pi T\int dyf^{\prime}[\theta(y)-\theta(-y)],\\
 \int d^{4-\epsilon}pf\delta^{\prime}(p^{2})
 & = & \frac{\pi^{\frac{3-\epsilon}{2}}T^{-\epsilon}}{2\Gamma\left(\frac{3-\epsilon}{2}\right)}\int dy(1-\epsilon)y^{-1-\epsilon}f\left[\theta(y)-\theta(y)\right],
\end{eqnarray}
and the zeroth-distribution function is
\begin{eqnarray}
f & = & \frac{1}{4\pi^{3}}\left[\frac{1}{e^{y-\bar{\mu}_{s}}+1}\theta(y)+\left(\frac{1}{e^{-y+\bar{\mu}_{s}}+1}-1\right)\theta(-y)\right].
\end{eqnarray}
Employing the integral formula
\begin{eqnarray}
\int_{0}^{\infty}dy\left[\frac{1}{e^{y-\bar{\mu}_{s}}+1}-\frac{1}{e^{y+\bar{\mu}_{s}}+1}\right] &=& \bar{\mu}_{s}
\end{eqnarray}
The convergence coefficient is given by
\begin{eqnarray}
\int d^{4}pp_uf^{\prime}\delta^{\prime}(p^{2}) &=& -\frac{1}{4\pi^{2}}T.
\end{eqnarray}
The divergence coefficient can be denoted as

\begin{eqnarray}
\int d^{4-\epsilon}pf\delta^{\prime}(p^{2}) &=& \frac{\left(1-\epsilon\right)}{4}\kappa^{\epsilon},
\end{eqnarray}
where
\begin{eqnarray}
\kappa_{s}^{\epsilon} & = & \frac{4\pi^{\frac{3-\epsilon}{2}}T^{-\epsilon}}{\Gamma\left(\frac{3-\epsilon}{2}\right)(2\pi)^{3-\epsilon}}\int dyy^{-1-\epsilon}\left(\frac{1}{e^{y-\bar{\mu}_{s}}+1}+\frac{1}{e^{y+\bar{\mu}_{s}}+1}-1\right).
\end{eqnarray}

\bibliographystyle{h-physrev}
\bibliography{WF-varying-EM}

\begin{thebibliography}{100}

\bibitem{Liang:2004ph}
Z.-T. Liang and X.-N. Wang,
\newblock Phys. Rev. Lett. {\bf 94}, 102301 (2005), nucl-th/0410079,
\newblock [Erratum: Phys.Rev.Lett. 96, 039901 (2006)].

\bibitem{Gao:2007bc}
J.-H. Gao {\em et~al.},
\newblock Phys. Rev. C {\bf 77}, 044902 (2008), 0710.2943.

\bibitem{Becattini:2007sr}
F.~Becattini, F.~Piccinini, and J.~Rizzo,
\newblock Phys. Rev. C {\bf 77}, 024906 (2008), 0711.1253.

\bibitem{Bzdak:2011yy}
A.~Bzdak and V.~Skokov,
\newblock Phys. Lett. B {\bf 710}, 171 (2012), 1111.1949.

\bibitem{Deng:2012pc}
W.-T. Deng and X.-G. Huang,
\newblock Phys. Rev. C {\bf 85}, 044907 (2012), 1201.5108.

\bibitem{Bloczynski:2012en}
J.~Bloczynski, X.-G. Huang, X.~Zhang, and J.~Liao,
\newblock Phys. Lett. {\bf B718}, 1529 (2013), 1209.6594.

\bibitem{Vilenkin:1980fu}
A.~Vilenkin,
\newblock Phys. Rev. {\bf D22}, 3080 (1980).

\bibitem{Kharzeev:2007jp}
D.~E. Kharzeev, L.~D. McLerran, and H.~J. Warringa,
\newblock Nucl. Phys. {\bf A803}, 227 (2008), 0711.0950.

\bibitem{Fukushima:2008xe}
K.~Fukushima, D.~E. Kharzeev, and H.~J. Warringa,
\newblock Phys. Rev. {\bf D78}, 074033 (2008), 0808.3382.

\bibitem{Vilenkin:1978hb}
A.~Vilenkin,
\newblock Phys. Lett. B {\bf 80}, 150 (1978).

\bibitem{Kharzeev:2007tn}
D.~Kharzeev and A.~Zhitnitsky,
\newblock Nucl. Phys. A {\bf 797}, 67 (2007), 0706.1026.

\bibitem{Erdmenger:2008rm}
J.~Erdmenger, M.~Haack, M.~Kaminski, and A.~Yarom,
\newblock JHEP {\bf 01}, 055 (2009), 0809.2488.

\bibitem{Banerjee:2008th}
N.~Banerjee {\em et~al.},
\newblock JHEP {\bf 01}, 094 (2011), 0809.2596.

\bibitem{Son:2004tq}
D.~T. Son and A.~R. Zhitnitsky,
\newblock Phys. Rev. D {\bf 70}, 074018 (2004), hep-ph/0405216.

\bibitem{Metlitski:2005pr}
M.~A. Metlitski and A.~R. Zhitnitsky,
\newblock Phys. Rev. D {\bf 72}, 045011 (2005), hep-ph/0505072.

\bibitem{Gao:2012ix}
J.-H. Gao, Z.-T. Liang, S.~Pu, Q.~Wang, and X.-N. Wang,
\newblock Phys. Rev. Lett. {\bf 109}, 232301 (2012), 1203.0725.

\bibitem{Newman:2005hd}
G.~M. Newman,
\newblock JHEP {\bf 01}, 158 (2006), hep-ph/0511236.

\bibitem{Yee:2009vw}
H.-U. Yee,
\newblock JHEP {\bf 11}, 085 (2009), 0908.4189.

\bibitem{Rebhan:2009vc}
A.~Rebhan, A.~Schmitt, and S.~A. Stricker,
\newblock JHEP {\bf 01}, 026 (2010), 0909.4782.

\bibitem{Gorsky:2010xu}
A.~Gorsky, P.~N. Kopnin, and A.~V. Zayakin,
\newblock Phys. Rev. D {\bf 83}, 014023 (2011), 1003.2293.

\bibitem{Gynther:2010ed}
A.~Gynther, K.~Landsteiner, F.~Pena-Benitez, and A.~Rebhan,
\newblock JHEP {\bf 02}, 110 (2011), 1005.2587.

\bibitem{Hoyos:2011us}
C.~Hoyos, T.~Nishioka, and A.~O'Bannon,
\newblock JHEP {\bf 10}, 084 (2011), 1106.4030.

\bibitem{Amado:2011zx}
I.~Amado, K.~Landsteiner, and F.~Pena-Benitez,
\newblock JHEP {\bf 05}, 081 (2011), 1102.4577.

\bibitem{Nair:2011mk}
V.~P. Nair, R.~Ray, and S.~Roy,
\newblock Phys. Rev. D {\bf 86}, 025012 (2012), 1112.4022.

\bibitem{Kalaydzhyan:2011vx}
T.~Kalaydzhyan and I.~Kirsch,
\newblock Phys. Rev. Lett. {\bf 106}, 211601 (2011), 1102.4334.

\bibitem{Lin:2013sga}
S.~Lin and H.-U. Yee,
\newblock Phys. Rev. D {\bf 88}, 025030 (2013), 1305.3949.

\bibitem{Pu:2014cwa}
S.~Pu, S.-Y. Wu, and D.-L. Yang,
\newblock Phys. Rev. {\bf D89}, 085024 (2014), 1401.6972.

\bibitem{Pu:2014fva}
S.~Pu, S.-Y. Wu, and D.-L. Yang,
\newblock Phys. Rev. {\bf D91}, 025011 (2015), 1407.3168.

\bibitem{Kharzeev:2009pj}
D.~E. Kharzeev and H.~J. Warringa,
\newblock Phys. Rev. D {\bf 80}, 034028 (2009), 0907.5007.

\bibitem{Fukushima:2009ft}
K.~Fukushima, D.~E. Kharzeev, and H.~J. Warringa,
\newblock Nucl. Phys. A {\bf 836}, 311 (2010), 0912.2961.

\bibitem{Asakawa:2010bu}
M.~Asakawa, A.~Majumder, and B.~Muller,
\newblock Phys. Rev. C {\bf 81}, 064912 (2010), 1003.2436.

\bibitem{Fukushima:2010vw}
K.~Fukushima, D.~E. Kharzeev, and H.~J. Warringa,
\newblock Phys. Rev. Lett. {\bf 104}, 212001 (2010), 1002.2495.

\bibitem{Fukushima:2010zza}
K.~Fukushima and M.~Ruggieri,
\newblock Phys. Rev. D {\bf 82}, 054001 (2010), 1004.2769.

\bibitem{Landsteiner:2011cp}
K.~Landsteiner, E.~Megias, and F.~Pena-Benitez,
\newblock Phys. Rev. Lett. {\bf 107}, 021601 (2011), 1103.5006.

\bibitem{Hou:2011ze}
D.~Hou, H.~Liu, and H.-c. Ren,
\newblock JHEP {\bf 05}, 046 (2011), 1103.2035.

\bibitem{Hou:2012xg}
D.-F. Hou, H.~Liu, and H.-c. Ren,
\newblock Phys. Rev. D {\bf 86}, 121703 (2012), 1210.0969.

\bibitem{Chen:2016xtg}
J.-W. Chen, T.~Ishii, S.~Pu, and N.~Yamamoto,
\newblock Phys. Rev. D {\bf 93}, 125023 (2016), 1603.03620.

\bibitem{Lin:2018aon}
S.~Lin and L.~Yang,
\newblock Phys. Rev. D {\bf 98}, 114022 (2018), 1810.02979.

\bibitem{Feng:2018tpb}
B.~Feng, D.-F. Hou, and H.-C. Ren,
\newblock Phys. Rev. D {\bf 99}, 036010 (2019), 1810.05954.

\bibitem{Dong:2020zci}
R.-D. Dong, R.-H. Fang, D.-F. Hou, and D.~She,
\newblock Chin. Phys. C {\bf 44}, 074106 (2020), 2001.05801.

\bibitem{Son:2009tf}
D.~T. Son and P.~Surowka,
\newblock Phys. Rev. Lett. {\bf 103}, 191601 (2009), 0906.5044.

\bibitem{Sadofyev:2010pr}
A.~V. Sadofyev and M.~V. Isachenkov,
\newblock Phys. Lett. B {\bf 697}, 404 (2011), 1010.1550.

\bibitem{Pu:2010as}
S.~Pu, J.-h. Gao, and Q.~Wang,
\newblock Phys. Rev. D {\bf 83}, 094017 (2011), 1008.2418.

\bibitem{Kharzeev:2011ds}
D.~E. Kharzeev and H.-U. Yee,
\newblock Phys. Rev. D {\bf 84}, 045025 (2011), 1105.6360.

\bibitem{Stephanov:2012ki}
M.~A. Stephanov and Y.~Yin,
\newblock Phys. Rev. Lett. {\bf 109}, 162001 (2012), 1207.0747.

\bibitem{Son:2012zy}
D.~T. Son and N.~Yamamoto,
\newblock Phys. Rev. D {\bf 87}, 085016 (2013), 1210.8158.

\bibitem{Chen:2012ca}
J.-W. Chen, S.~Pu, Q.~Wang, and X.-N. Wang,
\newblock Phys. Rev. Lett. {\bf 110}, 262301 (2013), 1210.8312.

\bibitem{Manuel:2013zaa}
C.~Manuel and J.~M. Torres-Rincon,
\newblock Phys. Rev. D {\bf 89}, 096002 (2014), 1312.1158.

\bibitem{Chen:2014cla}
J.-Y. Chen, D.~T. Son, M.~A. Stephanov, H.-U. Yee, and Y.~Yin,
\newblock Phys. Rev. Lett. {\bf 113}, 182302 (2014), 1404.5963.

\bibitem{Chen:2015gta}
J.-Y. Chen, D.~T. Son, and M.~A. Stephanov,
\newblock Phys. Rev. Lett. {\bf 115}, 021601 (2015), 1502.06966.

\bibitem{Hidaka:2016yjf}
Y.~Hidaka, S.~Pu, and D.-L. Yang,
\newblock Phys. Rev. D {\bf 95}, 091901 (2017), 1612.04630.

\bibitem{Mueller:2017lzw}
N.~Mueller and R.~Venugopalan,
\newblock Phys. Rev. D {\bf 97}, 051901 (2018), 1701.03331.

\bibitem{Huang:2017tsq}
A.~Huang, Y.~Jiang, S.~Shi, J.~Liao, and P.~Zhuang,
\newblock Phys. Lett. B {\bf 777}, 177 (2018), 1703.08856.

\bibitem{Huang:2018wdl}
A.~Huang, S.~Shi, Y.~Jiang, J.~Liao, and P.~Zhuang,
\newblock Phys. Rev. D {\bf 98}, 036010 (2018), 1801.03640.

\bibitem{Hidaka:2018ekt}
Y.~Hidaka and D.-L. Yang,
\newblock Phys. Rev. D {\bf 98}, 016012 (2018), 1801.08253.

\bibitem{Gao:2018wmr}
J.-H. Gao, Z.-T. Liang, Q.~Wang, and X.-N. Wang,
\newblock Phys. Rev. D {\bf 98}, 036019 (2018), 1802.06216.

\bibitem{Gao:2018jsi}
J.-h. Gao, J.-Y. Pang, and Q.~Wang,
\newblock Phys. Rev. D {\bf 100}, 016008 (2019), 1810.02028.

\bibitem{Liu:2018xip}
Y.-C. Liu, L.-L. Gao, K.~Mameda, and X.-G. Huang,
\newblock Phys. Rev. D {\bf 99}, 085014 (2019), 1812.10127.

\bibitem{STAR:2017ckg}
STAR, L.~Adamczyk {\em et~al.},
\newblock Nature {\bf 548}, 62 (2017), 1701.06657.

\bibitem{STAR:2019erd}
STAR, J.~Adam {\em et~al.},
\newblock Phys. Rev. Lett. {\bf 123}, 132301 (2019), 1905.11917.

\bibitem{ALICE:2019aid}
ALICE, S.~Acharya {\em et~al.},
\newblock Phys. Rev. Lett. {\bf 125}, 012301 (2020), 1910.14408.

\bibitem{STAR:2020xbm}
STAR, J.~Adam {\em et~al.},
\newblock Phys. Rev. Lett. {\bf 126}, 162301 (2021), 2012.13601.

\bibitem{STAR:2022fan}
STAR, M.~S. Abdallah {\em et~al.},
\newblock Nature {\bf 614}, 244 (2023), 2204.02302.

\bibitem{ALICE:2023jad}
ALICE, S.~Acharya {\em et~al.},
\newblock Phys. Rev. Lett. {\bf 131}, 042303 (2023).

\bibitem{Liang:2004xn}
Z.-T. Liang and X.-N. Wang,
\newblock Phys. Lett. B {\bf 629}, 20 (2005), nucl-th/0411101.

\bibitem{Karpenko:2016jyx}
I.~Karpenko and F.~Becattini,
\newblock Eur. Phys. J. C {\bf 77}, 213 (2017), 1610.04717.

\bibitem{Becattini:2017gcx}
F.~Becattini and I.~Karpenko,
\newblock Phys. Rev. Lett. {\bf 120}, 012302 (2018), 1707.07984.

\bibitem{Xie:2017upb}
Y.~Xie, D.~Wang, and L.~P. Csernai,
\newblock Phys. Rev. C {\bf 95}, 031901 (2017), 1703.03770.

\bibitem{Pang:2016igs}
L.-G. Pang, H.~Petersen, Q.~Wang, and X.-N. Wang,
\newblock Phys. Rev. Lett. {\bf 117}, 192301 (2016), 1605.04024.

\bibitem{Li:2017slc}
H.~Li, L.-G. Pang, Q.~Wang, and X.-L. Xia,
\newblock Phys. Rev. C {\bf 96}, 054908 (2017), 1704.01507.

\bibitem{Wei:2018zfb}
D.-X. Wei, W.-T. Deng, and X.-G. Huang,
\newblock Phys. Rev. C {\bf 99}, 014905 (2019), 1810.00151.

\bibitem{Ryu:2021lnx}
S.~Ryu, V.~Jupic, and C.~Shen,
\newblock Phys. Rev. C {\bf 104}, 054908 (2021), 2106.08125.

\bibitem{Shi:2017wpk}
S.~Shi, K.~Li, and J.~Liao,
\newblock Phys. Lett. B {\bf 788}, 409 (2019), 1712.00878.

\bibitem{Fu:2020oxj}
B.~Fu, K.~Xu, X.-G. Huang, and H.~Song,
\newblock Phys. Rev. C {\bf 103}, 024903 (2021), 2011.03740.

\bibitem{Sun:2017xhx}
Y.~Sun and C.~M. Ko,
\newblock Phys. Rev. {\bf C96}, 024906 (2017), 1706.09467.

\bibitem{Wu:2022mkr}
X.-Y. Wu, C.~Yi, G.-Y. Qin, and S.~Pu,
\newblock Phys. Rev. C {\bf 105}, 064909 (2022), 2204.02218.

\bibitem{Alzhrani:2022dpi}
S.~Alzhrani, S.~Ryu, and C.~Shen,
\newblock Phys. Rev. C {\bf 106}, 014905 (2022), 2203.15718.

\bibitem{Liu:2021uhn}
S.~Y.~F. Liu and Y.~Yin,
\newblock JHEP {\bf 07}, 188 (2021), 2103.09200.

\bibitem{Fu:2021pok}
B.~Fu, S.~Y.~F. Liu, L.~Pang, H.~Song, and Y.~Yin,
\newblock Phys. Rev. Lett. {\bf 127}, 142301 (2021), 2103.10403.

\bibitem{Becattini:2021suc}
F.~Becattini, M.~Buzzegoli, and A.~Palermo,
\newblock Phys. Lett. B {\bf 820}, 136519 (2021), 2103.10917.

\bibitem{Becattini:2021iol}
F.~Becattini, M.~Buzzegoli, G.~Inghirami, I.~Karpenko, and A.~Palermo,
\newblock Phys. Rev. Lett. {\bf 127}, 272302 (2021), 2103.14621.

\bibitem{Hidaka:2017auj}
Y.~Hidaka, S.~Pu, and D.-L. Yang,
\newblock Phys. Rev. D {\bf 97}, 016004 (2018), 1710.00278.

\bibitem{Yi:2021ryh}
C.~Yi, S.~Pu, and D.-L. Yang,
\newblock Phys. Rev. C {\bf 104}, 064901 (2021), 2106.00238.

\bibitem{Fu:2022myl}
B.~Fu, L.~Pang, H.~Song, and Y.~Yin,
\newblock (2022), 2201.12970.

\bibitem{Weickgenannt:2020aaf}
N.~Weickgenannt, E.~Speranza, X.-l. Sheng, Q.~Wang, and D.~H. Rischke,
\newblock Phys. Rev. Lett. {\bf 127}, 052301 (2021), 2005.01506.

\bibitem{Yang:2020hri}
D.-L. Yang, K.~Hattori, and Y.~Hidaka,
\newblock JHEP {\bf 07}, 070 (2020), 2002.02612.

\bibitem{Weickgenannt:2021cuo}
N.~Weickgenannt, E.~Speranza, X.-l. Sheng, Q.~Wang, and D.~H. Rischke,
\newblock Phys. Rev. D {\bf 104}, 016022 (2021), 2103.04896.

\bibitem{Sheng:2021kfc}
X.-L. Sheng, N.~Weickgenannt, E.~Speranza, D.~H. Rischke, and Q.~Wang,
\newblock Phys. Rev. D {\bf 104}, 016029 (2021), 2103.10636.

\bibitem{Wang:2020pej}
Z.~Wang, X.~Guo, and P.~Zhuang,
\newblock Eur. Phys. J. C {\bf 81}, 799 (2021), 2009.10930.

\bibitem{Wang:2021qnt}
Z.~Wang and P.~Zhuang,
\newblock (2021), 2105.00915.

\bibitem{Weickgenannt:2022jes}
N.~Weickgenannt, D.~Wagner, and E.~Speranza,
\newblock Phys. Rev. D {\bf 105}, 116026 (2022), 2204.01797.

\bibitem{Lin:2021mvw}
S.~Lin,
\newblock Phys. Rev. D {\bf 105}, 076017 (2022), 2109.00184.

\bibitem{Fang:2022ttm}
S.~Fang, S.~Pu, and D.-L. Yang,
\newblock Phys. Rev. D {\bf 106}, 016002 (2022), 2204.11519.

\bibitem{Wang:2022yli}
Z.~Wang,
\newblock Phys. Rev. D {\bf 106}, 076011 (2022), 2205.09334.

\bibitem{Lin:2022tma}
S.~Lin and Z.~Wang,
\newblock JHEP {\bf 12}, 030 (2022), 2206.12573.

\bibitem{Li:2019qkf}
S.~Li and H.-U. Yee,
\newblock Phys. Rev. D {\bf 100}, 056022 (2019), 1905.10463.

\bibitem{Wagner:2022amr}
D.~Wagner, N.~Weickgenannt, and D.~H. Rischke,
\newblock Phys. Rev. D {\bf 106}, 116021 (2022), 2210.06187.

\bibitem{Yamamoto:2023okm}
N.~Yamamoto and D.-L. Yang,
\newblock (2023), 2308.08257.

\bibitem{Fang:2023bbw}
S.~Fang, S.~Pu, and D.-L. Yang,
\newblock Phys. Rev. D {\bf 109}, 034034 (2024), 2311.15197.

\bibitem{Lin:2024zik}
S.~Lin and Z.~Wang,
\newblock (2024), 2406.10003.

\bibitem{Fang:2024vds}
S.~Fang and S.~Pu,
\newblock (2024), 2408.09877.

\bibitem{STAR:2023eck}
STAR, M.~Abdulhamid {\em et~al.},
\newblock Phys. Rev. Lett. {\bf 131}, 202301 (2023), 2303.09074.

\bibitem{Yi:2024kwu}
C.~Yi, X.-Y. Wu, J.~Zhu, S.~Pu, and G.-Y. Qin,
\newblock (2024), 2408.04296.

\bibitem{Becattini:2020ngo}
F.~Becattini and M.~A. Lisa,
\newblock Ann. Rev. Nucl. Part. Sci. {\bf 70}, 395 (2020), 2003.03640.

\bibitem{Gao:2020vbh}
J.-H. Gao, G.-L. Ma, S.~Pu, and Q.~Wang,
\newblock Nucl. Sci. Tech. {\bf 31}, 90 (2020), 2005.10432.

\bibitem{Hidaka:2022dmn}
Y.~Hidaka, S.~Pu, Q.~Wang, and D.-L. Yang,
\newblock Prog. Part. Nucl. Phys. {\bf 127}, 103989 (2022), 2201.07644.

\bibitem{Becattini:2022zvf}
F.~Becattini,
\newblock Rept. Prog. Phys. {\bf 85}, 122301 (2022), 2204.01144.

\bibitem{Becattini:2024uha}
F.~Becattini {\em et~al.},
\newblock (2024), 2402.04540.

\bibitem{Roy:2015kma}
V.~Roy, S.~Pu, L.~Rezzolla, and D.~Rischke,
\newblock Phys. Lett. {\bf B750}, 45 (2015), 1506.06620.

\bibitem{Roy:2015coa}
V.~Roy and S.~Pu,
\newblock Phys. Rev. {\bf C92}, 064902 (2015), 1508.03761.

\bibitem{Pu:2016bxy}
S.~Pu and D.-L. Yang,
\newblock Phys. Rev. {\bf D93}, 054042 (2016), 1602.04954.

\bibitem{Inghirami:2016iru}
G.~Inghirami {\em et~al.},
\newblock Eur. Phys. J. {\bf C76}, 659 (2016), 1609.03042.

\bibitem{Pu:2016ayh}
S.~Pu, V.~Roy, L.~Rezzolla, and D.~H. Rischke,
\newblock Phys. Rev. {\bf D93}, 074022 (2016), 1602.04953.

\bibitem{Pu:2016rdq}
S.~Pu and D.-L. Yang,
\newblock EPJ Web Conf. {\bf 137}, 13021 (2017), 1611.04840.

\bibitem{Siddique:2019gqh}
I.~Siddique, R.-j. Wang, S.~Pu, and Q.~Wang,
\newblock Phys. Rev. {\bf D99}, 114029 (2019), 1904.01807.

\bibitem{Peng:2022cya}
H.-H. Peng, S.~Wu, R.-j. Wang, D.~She, and S.~Pu,
\newblock Phys. Rev. D {\bf 107}, 096010 (2023), 2211.11286.

\bibitem{Wang:2021oqq}
Z.~Wang, J.~Zhao, C.~Greiner, Z.~Xu, and P.~Zhuang,
\newblock Phys. Rev. C {\bf 105}, L041901 (2022), 2110.14302.

\bibitem{Yan:2021zjc}
L.~Yan and X.-G. Huang,
\newblock (2021), 2104.00831.

\bibitem{Zhang:2022lje}
J.-J. Zhang {\em et~al.},
\newblock Phys. Rev. Res. {\bf 4}, 033138 (2022), 2201.06171.

\bibitem{Banerjee:2012iz}
N.~Banerjee {\em et~al.},
\newblock JHEP {\bf 09}, 046 (2012), 1203.3544.

\bibitem{Bhattacharyya:2013ida}
S.~Bhattacharyya, J.~R. David, and S.~Thakur,
\newblock JHEP {\bf 01}, 010 (2014), 1305.0340.

\bibitem{Megias:2014mba}
E.~Megias and M.~Valle,
\newblock JHEP {\bf 11}, 005 (2014), 1408.0165.

\bibitem{Bu:2019qmd}
Y.~Bu and S.~Lin,
\newblock Eur. Phys. J. C {\bf 80}, 401 (2020), 1912.11277.

\bibitem{Jimenez-Alba:2015bia}
A.~Jimenez-Alba and H.-U. Yee,
\newblock Phys. Rev. D {\bf 92}, 014023 (2015), 1504.05866.

\bibitem{Hattori:2016njk}
K.~Hattori and Y.~Yin,
\newblock Phys. Rev. Lett. {\bf 117}, 152002 (2016), 1607.01513.

\bibitem{Buzzegoli:2017cqy}
M.~Buzzegoli, E.~Grossi, and F.~Becattini,
\newblock JHEP {\bf 10}, 091 (2017), 1704.02808,
\newblock [Erratum: JHEP 07, 119 (2018)].

\bibitem{Buzzegoli:2018wpy}
M.~Buzzegoli and F.~Becattini,
\newblock JHEP {\bf 12}, 002 (2018), 1807.02071,
\newblock [Erratum: JHEP 03, 045 (2022)].

\bibitem{Becattini:2020qol}
F.~Becattini, M.~Buzzegoli, and A.~Palermo,
\newblock JHEP {\bf 02}, 101 (2021), 2007.08249.

\bibitem{Palermo:2021hlf}
A.~Palermo, M.~Buzzegoli, and F.~Becattini,
\newblock JHEP {\bf 10}, 077 (2021), 2106.08340.

\bibitem{Satow:2014lia}
D.~Satow,
\newblock Phys. Rev. D {\bf 90}, 034018 (2014), 1406.7032.

\bibitem{Gorbar:2017cwv}
E.~V. Gorbar, V.~A. Miransky, I.~A. Shovkovy, and P.~O. Sukhachov,
\newblock Phys. Rev. B {\bf 95}, 205141 (2017), 1702.02950.

\bibitem{Gorbar:2017toh}
E.~V. Gorbar, D.~O. Rybalka, and I.~A. Shovkovy,
\newblock Phys. Rev. D {\bf 95}, 096010 (2017), 1702.07791.

\bibitem{Mameda:2023ueq}
K.~Mameda,
\newblock Phys. Rev. D {\bf 108}, 016001 (2023), 2305.02134.

\bibitem{Muller:2018ibh}
B.~M\"uller and A.~Sch\"afer,
\newblock Phys. Rev. D {\bf 98}, 071902 (2018), 1806.10907.

\bibitem{Guo:2019joy}
Y.~Guo, S.~Shi, S.~Feng, and J.~Liao,
\newblock Phys. Lett. B {\bf 798}, 134929 (2019), 1905.12613.

\bibitem{Buzzegoli:2022qrr}
M.~Buzzegoli,
\newblock Nucl. Phys. A {\bf 1036}, 122674 (2023), 2211.04549.

\bibitem{Xu:2022hql}
K.~Xu, F.~Lin, A.~Huang, and M.~Huang,
\newblock Phys. Rev. D {\bf 106}, L071502 (2022), 2205.02420.

\bibitem{Yang:2020mtz}
S.-Z. Yang, J.-H. Gao, Z.-T. Liang, and Q.~Wang,
\newblock Phys. Rev. D {\bf 102}, 116024 (2020), 2003.04517.

\bibitem{Yi:2021unq}
C.~Yi, S.~Pu, J.-H. Gao, and D.-L. Yang,
\newblock (2021), 2112.15531.

\bibitem{Yi:2023tgg}
C.~Yi {\em et~al.},
\newblock (2023), 2304.08777.

\bibitem{Vasak:1987um}
D.~Vasak, M.~Gyulassy, and H.~T. Elze,
\newblock Annals Phys. {\bf 173}, 462 (1987).

\bibitem{Gao:2019zhk}
J.-H. Gao, Z.-T. Liang, and Q.~Wang,
\newblock Phys. Rev. D {\bf 101}, 096015 (2020), 1910.11060.

\bibitem{Becattini:2013fla}
F.~Becattini, V.~Chandra, L.~Del~Zanna, and E.~Grossi,
\newblock Annals Phys. {\bf 338}, 32 (2013), 1303.3431.

\bibitem{Fang:2016uds}
R.-h. Fang, J.-y. Pang, Q.~Wang, and X.-n. Wang,
\newblock Phys. Rev. D {\bf 95}, 014032 (2017), 1611.04670.

\bibitem{Gao:2021rom}
J.-H. Gao,
\newblock Phys. Rev. D {\bf 104}, 076016 (2021), 2105.08293.

\bibitem{Hattori:2019lfp}
K.~Hattori, M.~Hongo, X.-G. Huang, M.~Matsuo, and H.~Taya,
\newblock Phys. Lett. B {\bf 795}, 100 (2019), 1901.06615.

\bibitem{Fukushima:2020ucl}
K.~Fukushima and S.~Pu,
\newblock Phys. Lett. B {\bf 817}, 136346 (2021), 2010.01608.

\bibitem{Hongo:2021ona}
M.~Hongo, X.-G. Huang, M.~Kaminski, M.~Stephanov, and H.-U. Yee,
\newblock JHEP {\bf 11}, 150 (2021), 2107.14231.

\bibitem{Daher:2022wzf}
A.~Daher, A.~Das, and R.~Ryblewski,
\newblock Phys. Rev. D {\bf 107}, 054043 (2023), 2209.10460.

\bibitem{Xie:2023gbo}
X.-Q. Xie, D.-L. Wang, C.~Yang, and S.~Pu,
\newblock (2023), 2306.13880.

\bibitem{Ren:2024pur}
X.~Ren, C.~Yang, D.-L. Wang, and S.~Pu,
\newblock Phys. Rev. D {\bf 110}, 034010 (2024), 2405.03105.

\bibitem{Wang:2021ngp}
D.-L. Wang, S.~Fang, and S.~Pu,
\newblock Phys. Rev. D {\bf 104}, 114043 (2021), 2107.11726.

\bibitem{Wang:2021wqq}
D.-L. Wang, X.-Q. Xie, S.~Fang, and S.~Pu,
\newblock Phys. Rev. D {\bf 105}, 114050 (2022), 2112.15535.

\bibitem{Wang:2024afv}
D.-L. Wang, L.~Yan, and S.~Pu,
\newblock (2024), 2408.03781.

\end{thebibliography}

\end{document}